 \documentclass[10pt, final]{IEEEtran}

\usepackage{amsthm}

\usepackage{amssymb}
\usepackage{amsmath}
\usepackage{bbm}
\usepackage[ruled]{algorithm2e}
\usepackage{mathrsfs}
\usepackage{cite}
\usepackage{bm,epsfig,amsthm,url}
\usepackage{indentfirst}
\usepackage{amsfonts}
\usepackage{epstopdf}
\usepackage{cases}
\usepackage[caption=false,font=scriptsize]{subfig}
\usepackage{xcolor}
\usepackage{mathtools}

\makeatletter
\renewcommand{\maketag@@@}[1]{\hbox{\m@th\normalsize\normalfont#1}}%
\makeatother
\usepackage{hyperref}
\usepackage{stfloats}

\newtheoremstyle{mystyle}{}{}{}{}{}{: }{0pt}{\indent \it{\thmname{#1}\thmnumber{ #2}\thmnote{#3}}}
\theoremstyle{mystyle}

\newtheorem{Proposition}{Proposition}

\newtheorem{Remark}{Remark}

\begin{document}

\title{\huge Joint Size and Placement Optimization for IRS-Aided Communications with Active and Passive Elements}

\author{{Qiaoyan~Peng,~Qingqing~Wu,~Wen~Chen,~Chaoying~Huang,~Beixiong~Zheng,~Shaodan~Ma,~Mengnan~Jian,\\Yijian~Chen,~and~Jun~Yang}
\thanks{Q. Peng is with the State Key Laboratory of Internet of Things for Smart City, University of Macau, Macao 999078, China, and also with the Department of Electronic Engineering, Shanghai Jiao Tong University, Shanghai, 200240, China (email: qiaoyan.peng@connect.um.edu.mo). Q. Wu, W. Chen, and C. Huang are with the Department of Electronic Engineering, Shanghai Jiao Tong University, Shanghai 200240, China (e-mail: qingqingwu@sjtu.edu.cn; wenchen@sjtu.edu.cn; chaoyinghuang@sjtu.edu.cn;). B. Zheng is with the School of Microelectronics, South China University of Technology, Guangzhou 511442, China, and also with the Shenzhen Research Institute of Big Data, Shenzhen 518172, China (e-mail: bxzheng@scut.edu.cn). S. Ma is with the State Key Laboratory of Internet of Things for Smart City, University of Macau, Macao 999078, China (email: shaodanma@um.edu.mo). M. Jian, Y. Chen, and J. Yang are with ZTE Corporation, Shenzhen 518057, China (e-mail: jian.mengnan@zte.com.cn; chen.yijian@zte.com.cn; yang.jun10@zte.com.cn).
}
}

\maketitle

\begin{abstract}
Different types of intelligent reflecting surfaces (IRS) are exploited for assisting wireless communications. The joint use of passive IRS (PIRS) and active IRS (AIRS) emerges as a promising solution owing to their complementary advantages. They can be integrated into a single hybrid active-passive IRS (HIRS) or deployed in a distributed manner, which poses challenges in determining the IRS elements allocation and placement for rate maximization. In this paper, we investigate the capacity of an IRS-aided wireless communication system with both active and passive elements. Specifically, we consider three deployment schemes: 1) base station (BS)→HIRS→user (BHU); 2) BS→AIRS→PIRS→user (BAPU); 3) BS→PIRS→AIRS→user (BPAU). Under the line-of-sight channel model, we formulate a rate maximization problem via a joint optimization of the IRS elements allocation and placement. We first derive the optimized number of active and passive elements for BHU, BAPU, and BPAU schemes, respectively. Then, low-complexity HIRS/AIRS placement strategies are provided. To obtain more insights, we characterize the system capacity scaling orders for the three schemes with respect to the large total number of IRS elements, amplification power budget, and BS transmit power. Finally, simulation results are presented to validate our theoretical findings and show the performance difference among the BHU, BAPU, and BPAU schemes with the proposed joint design under various system setups.
\end{abstract}

\section{Introduction}
Intelligent reflecting surface (IRS) has gained significant attention in facilitating the widespread application of wireless networks due to its energy and cost efficiency as well as flexible deployment \cite{survey,BF,BF_DFRC,BF_ISAC,deployment}. Specifically, passive IRS (PIRS) can construct an intelligent wireless propagation environment by smartly adjusting the phase shift of each reflecting element \cite{singlePassive,singlePassive2,singlePassive3,zteirs1,zteirs2,zteirs3}. However, it suffers from severe product-distance path loss in practice. To overcome this challenge, active IRS (AIRS) has been proposed to enhance the reflected signal power thanks to its amplification capabilities, whereas it introduces negligible amplification noise \cite{singleActive,singleActive4,singleActive5,singleActive6}. 

Recently, extensive works have explored integrating AIRS and PIRS into a single system to leverage their complementary advantages. One of the effective ways is co-locating both active and passive elements to form a single large IRS, which is known as a hybrid active-passive IRS (HIRS) \cite{HIRS_SE,HIRS_EE,HIRS_MU,HIRS_Secure,HIRS_ISAC,HIRS,HIRS2,SPS1,SPS2}. Specifically, \cite{HIRS_MU} proposed alternating optimization (AO) based optimization algorithms for energy efficiency maximization in an HIRS-assisted downlink communication. The joint beamforming design was investigated in \cite{HIRS_Secure,HIRS_ISAC} for different systems, i.e., secure communication and millimeter-wave integrated sensing and communication systems, respectively. Benefiting from flexibly balancing the amplification gain and beamforming gain, the results in \cite{HIRS_SE,HIRS_EE,HIRS_MU,HIRS_Secure,HIRS_ISAC} demonstrated that the HIRS can outperform both AIRS and PIRS in most practical scenarios. In contrast, another design integrating AIRS and PIRS is deploying distributed AIRS and PIRS by grouping reflecting elements of the same type on two or more small IRS. Despite a higher power scaling order provided by the systems with two or more PIRSs thanks to the benefits of double or multiple reflections, multiplicative fading and beam routing remains critical issues in these systems \cite{doublePassive,doublePassive2,doublePassive3,routing1,routing2,routing3,survey_x2,routing4}. The AIRSs can be deployed to achieve signal amplification \cite{doubleActive,doubleActive2,doubleActive3}, whereas it involves more complex design considerations. This is because they require the joint optimization of both the reflection amplitude and phase of the active elements. In addition, a new design trade-off emerges between the received signal power maximization and the noise amplification power minimization. By taking advantage of AIRS and PIRS, \cite{AP_x,AP_N} studied the achievable rate maximization problem in two distributed AIRS and PIRS jointly aided communication systems, i.e., base station (BS)→AIRS→PIRS→user (BAPU) and BS→PIRS→AIRS→user (BPAU).

PIRS can achieve a higher beamforming gain when the number of reflecting elements is large, whereas AIRS can offer unique power amplification gain, resulting in significantly higher SNR with fewer reflecting elements \cite{HIRS_SE}. To fully reap the potential advantages of double IRS or HIRS to balance the trade-off between the amplification and beamforming gains, it is crucial to carefully design its elements allocation \cite{doublePIRS_N,HIRS_SE,HIRS_EE,HIRS_MEC,AP_N}. For example, \cite{doublePIRS_N} considered the scenario where two distributed PIRSs are respectively equipped with $N/2$ passive elements, which yields a power scaling order of $\mathcal{O}(N^4)$. For a communication system with the BHU scheme, i.e., BS→HIRS→user, \cite{HIRS_SE} focused on the ergodic capacity maximization, while considering constraints such as the total deployment and amplification power budget. \cite{HIRS_EE} further studied the impact of the number of active and passive elements at the HIRS on the system energy efficiency maximization. Moreover, the authors in \cite{HIRS_MEC} studied the HIRS elements allocation for minimizing the latency and energy consumption in a mobile edge computing system. With the joint use of AIRS and PIRS, \cite{AP_N} showed that the optimal number of passive elements should be twice that of active elements to maximize the achievable rate under two different deployment orders with a given deployment budget.

To fully exploit the superior flexibility and compatibility of the IRS for practical implementation, the IRS placement should be investigated in addition to its elements allocation design. The objective of optimizing IRS deployment is to enable the reflected signal to bypass obstacles, ensuring signal coverage and maximizing the received signal power, thereby further improving system performance \cite{survey_x1,multiPassive_x,multiPassive_x2,singleActive2,singleActive3,AP_x,survey_x2,x_BS,x_multi}. The results in \cite{survey_x1} demonstrated that the PIRS should be placed close to the BS or user to minimize the double path loss effect for a single-user setup. A double PIRS placement optimization problem was solved to mitigate inter-cell interference in air-ground communication networks in \cite{multiPassive_x}. The work was further extended to a multiple PIRS-aided multi-user communication system for the minimum achievable rate maximization in \cite{multiPassive_x2}. It was revealed in \cite{singleActive2,singleActive3} that the placement strategy of an AIRS is different from that of a PIRS to strike a balance between signal and noise amplification. In \cite{singleActive2}, the impact of the amplification power on the optimal location of AIRS was characterized, which showed that the amplification noise can be compensated by deploying the AIRS closer to the receiver. Moreover, \cite{AP_x} focused on the AIRS deployment design under BAPU and BPAU schemes. To summarize, the performance of the IRS-aided system with both active and passive elements is sensitive to the IRS location.

All the above works focus only on either IRS elements allocation or placement, but they affect each other. As such, it requires joint design, which takes into account the interaction between the two key factors. Under a more comprehensive consideration of system requirements, the joint elements allocation and placement design can be exploited to find a better strategy to improve system performance. Due to the above benefits, the authors in \cite{multiPassive_xN,doubleAIRS_xN,doubleAIRS_xN2} aimed to further improve the system's performance by jointly optimizing the IRS elements allocation and placement. For example, the authors of \cite{multiPassive_xN} studied the minimum outage probability with the joint design of the PIRS location and size optimization under various constraints. For a double AIRS-aided communication system, an efficient algorithm was proposed in \cite{doubleAIRS_xN} to solve the minimum rate maximization problem involving combinatorial optimization with a fixed per-element amplification power. For a single-user setup, \cite{doubleAIRS_xN2} derived the optimal solution in closed form for maximizing the achievable rate and obtained more insights under the total amplification power constraint. Regarding a system where both active and passive elements are deployed, several issues remain to be resolved. First, how to design the joint optimization of IRS elements allocation and placement? For BAPU and BPAU schemes, it can lead to performance degradation or even complete failure of the entire link if the performance of one of the IRSs is limited. This is due to the fact that signal transmission needs to pass through both AIRS and PIRS under both schemes. In contrast, the other type can compensate for the failure if one type of reflecting element at the HIRS fails, ensuring the signal transmission despite performance loss, which shows the robustness and flexibility of BHU. Given the different deployment strategies of AIRS and PIRS, it still remains unknown how to design the HIRS deployment for further improving the system rate performance. Second, do BAPU and BPAU always outperform BHU? This question arises from the cases of double AIRS and double PIRS compared to their single counterparts. BAPU and BPAU schemes may reap greater beamforming gains through cooperation because the double reflection of the signal results in higher signal power being received at the user and extends the coverage. 

\begin{figure*}[ht]
	\centering
	\subfloat[BHU scheme]{\label{fig:hybrid}\includegraphics[width=0.33\textwidth]{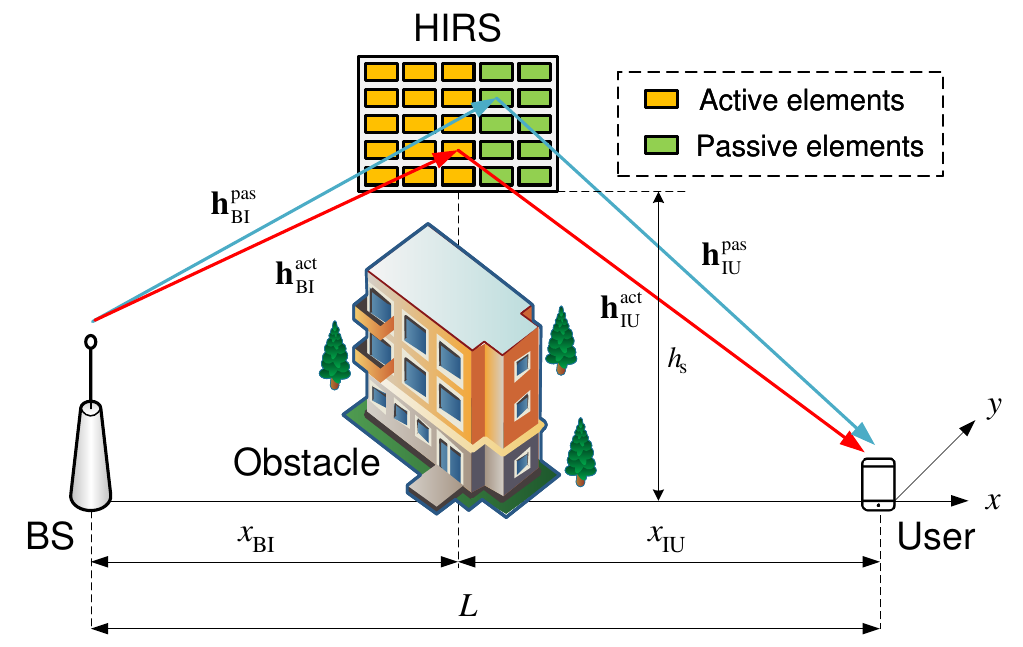}}
	\subfloat[BAPU scheme]{\label{fig:double_AP}\includegraphics[width=0.33\textwidth]{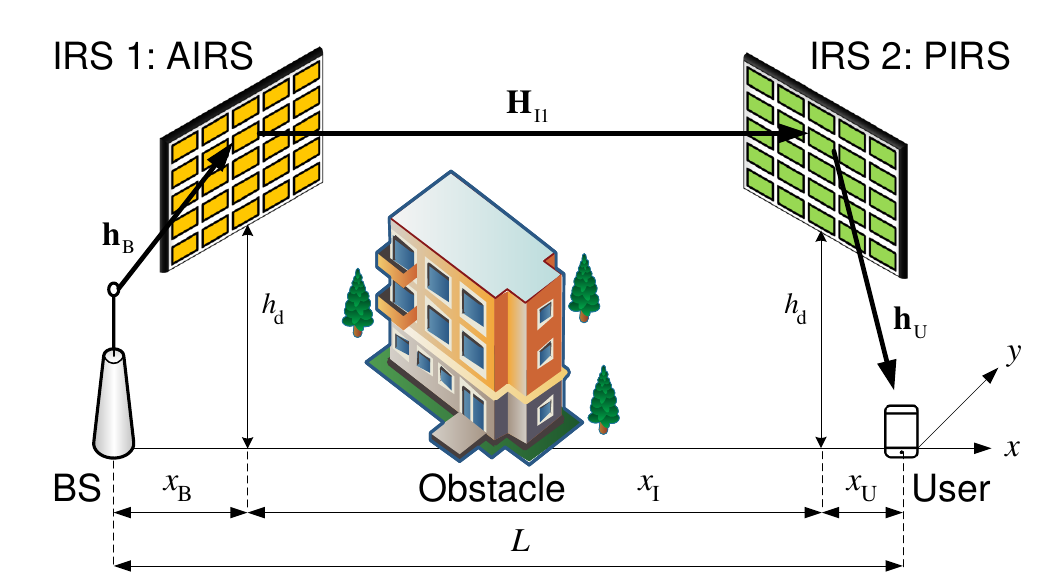}}
	\subfloat[BPAU scheme]{\label{fig:double_PA}\includegraphics[width=0.33\textwidth]{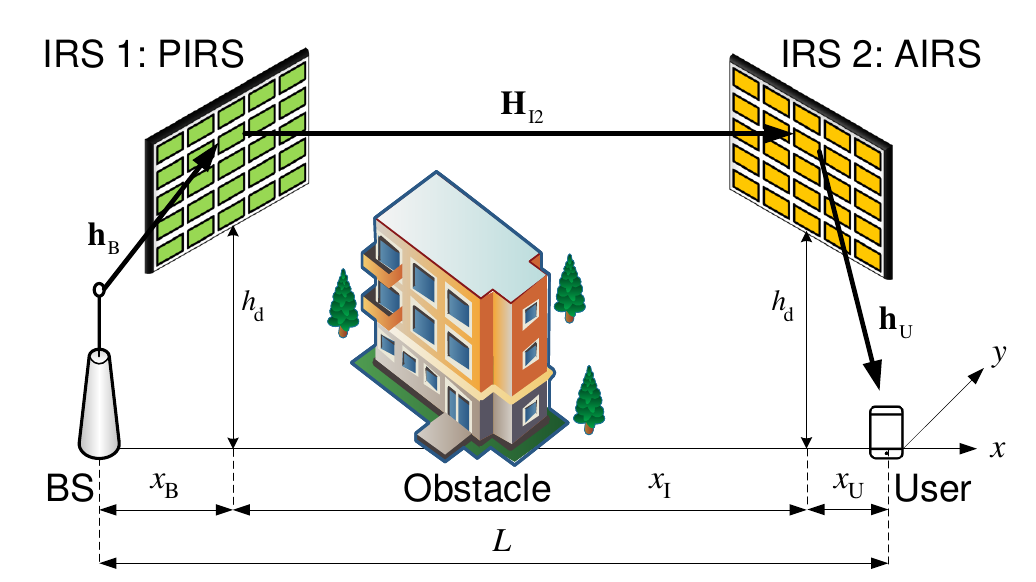}}
	\caption{Illustration of the considered systems.}
	\label{fig:systemmodel}
	\vspace{-10pt}
\end{figure*}

Motivated by the above, we focus on characterizing the capacity of the joint active and passive IRS-aided wireless system. \footnote{Note that the considered IRS architectures only have reflection capabilities, different from the intelligent omni-surface \cite{R4} and simultaneous transmitting and reflecting IRS \cite{R3}, which can achieve both signal reflection and refraction.} We aim to establish an analytical framework that enables a theoretical comparison of the capacity among three different deployment schemes, namely, BHU, BAPU, and BPAU (See Fig. \ref{fig:systemmodel}). The capacity characterization problem involves the joint design of the IRS beamforming, elements allocation, and placement, which renders it more challenging than the previous work. The main contributions of this paper are summarized as follows.
\begin{itemize}
\item First, we study a wireless communication system by comparing BHU, BAPU, and BPAU schemes. For the BHU scheme, a single-reflection link is provided via the HIRS, while both the BAPU and BPAU schemes establish a double-reflection link using two IRSs between the BS and the user. Based on the proposed models, the corresponding achievable rate maximization problems are formulated by jointly optimizing the active and passive elements allocation and IRS location.
\item Second, we derive the optimal number of active and passive elements for the three schemes. To obtain useful insights, we propose low-complexity solutions for BAPU and BPAU schemes in closed form with the given assumptions. Our analytic results show that more passive elements should be deployed at the PIRS for both BAPU and BPAU schemes as the total number of IRS elements $N$ increases. However, for the BHU scheme, the number of passive elements first remains at one and then linearly increases with respect to (w.r.t.) $N$ after it exceeds a threshold. Moreover, the BHU scheme is preferable when deploying the HIRS close to the user and the BAPU (BPAU) scheme performs best when the PIRS is close to the user (BS), which motivates us to further design the IRS placement. 
\item Based on the closed-form expressions of the IRS placement, we characterize the impact of key system parameters on the optimized horizontal BS-HIRS/AIRS distance. It reveals that the corresponding distance monotonically increases with the BS transmit power and decreases with the amplification power budget under the three schemes. Moreover, the results suggest that BPAU and BPAU should be avoided when the amplification power budget or transmit power is small, respectively.
\item To obtain more insights, we theoretically analyze the asymptotic SNR and then characterize the capacity scaling orders for the three schemes w.r.t. the total number of reflecting elements, transmit power, and amplification power, which lay the foundation for performance comparison. The results show that the BHU scheme outperforms both BAPU and BPAU schemes with a sufficient total number of IRS elements or transmit power, whereas the latter two schemes perform better when a large amplification power is affordable.
\item Finally, numerical results demonstrate that BHU is an energy-efficient solution for maximizing the capacity, especially in space-limited scenarios. In contrast, BAPU can achieve better rate performance when the transmit power budget is sufficiently large, whereas BPAU is preferable when the total number of IRS reflecting elements is large. Moreover, the results demonstrate that the joint active and passive IRS-aided wireless system with our proposed design can reap considerable performance gains over other benchmark systems.
\end{itemize}

The rest of this paper is organized as follows. Section \ref{systemModel} describes our system model and problem formulations considering BHU, BAPU, and BPAU schemes. Section \ref{EAP_opt} addresses the IRS elements allocation and placement problem. In Section \ref{order}, we analytically characterize the system capacity scaling orders. Section \ref{Simulation} presents simulation results to validate our theoretical analysis and provide further numerical comparison among the three schemes. Finally, we conclude in Section \ref{Conclusion}.

{\it{Notations:}} $\mathbb{N}^{^+}$ denotes the the set of positive natural numbers. For a complex-valued vector $\mathbf{x}$, ${\left[\mathbf{x}\right]}_m$ denotes the $m$-th entry of $\mathbf{x}$, $\arg \left( \mathbf{x} \right)$ represents its phase, and $\operatorname{diag} \left(\mathbf{x}\right)$ denotes a diagonal matrix with the elements in $\mathbf{x}$ on its main diagonal. $\mathbf{I}_{n}$ is an identity matrix of dimensions $n \times n$. $\left( \cdot \right) ^H$, $\frac{{\partial}}{{\partial }} \left( \cdot \right)$, and $\|\cdot\|$ stand for conjugate transpose operators, the partial derivative, and Euclidean norm, respectively. $\mathcal{CN} ( \mu,\sigma^2 )$ denotes the circularly symmetric complex Gaussian distribution with a mean of $\mu$ and variance of $\sigma^2$. $\mathcal{O} \left( \cdot \right)$ represents the big-O notation. $\otimes$ denotes the Kronecker product.

\section{System Model and Problem Formulations}
\label{systemModel}
As illustrated in Fig. \ref{fig:systemmodel}, we consider an IRS-aided wireless system, where the IRS with $N_\mathrm{p}$ passive elements and $N_\mathrm{a}$ active elements are deployed to enhance the transmission from a single-antenna BS to a single-antenna user \footnote{A typical user is selected to represent the performance of users in a quasi-static scenario, e.g., a hotspot area. This paper can be extended to multi-user systems by alternately optimizing the IRS reflection, elements allocation, and placement using similar optimization methods in \cite{doubleAIRS_xN} for a time-division multiple access scheme.} in a given zone. The direct BS-user, BS-IRS 2, and IRS 1-user links are assumed to be blocked due to dense obstacles, and the line-of-sight (LoS) channel model is used for other available links. According to \cite{CSI}, we assume that the channel state information of the considered links is available. Three different deployment schemes are considered, namely, BHU, BAPU, and BPAU. Specifically, a total of $N = N_\mathrm{p} + N_\mathrm{a}$ IRS elements form one single HIRS for the BHU scheme, which is deployed in the LoS region of the BS and the user (see Fig. \ref{fig:systemmodel} (a)). By contrast, passive and active elements are deployed at two distributed IRS for BAPU and BPAU schemes, where IRS 1 and IRS 2 are located in the vicinity of the BS and the user, respectively (see Fig. \ref{fig:systemmodel} (b) and (c)). Under a three-dimensional Cartesian coordinate system, the BS, the HIRS, IRS 1, IRS 2, and the user are located at $\boldsymbol{u}_\mathrm{B} = (0,0,0)$ and $\boldsymbol{u}_0 = (x_\mathrm{BI},0,h_\mathrm{s})$, $\boldsymbol{u}_1 = (x_\mathrm{B},0,h_\mathrm{d})$, $\boldsymbol{u}_2 = (L-x_\mathrm{U},0,h_\mathrm{d})$, and $\boldsymbol{u}_\mathrm{u} = (L,0,0)$, respectively.\footnote{Our analysis is applicable to IRS at any given height. For simplicity, we focus on the IRS position optimization in the horizontal direction. Nevertheless, the results can be extended to 3D spatial scenarios, where IRS may be located at different heights to avoid obstacles.} As such, the distances between the BS and HIRS, the HIRS and user, the BS and IRS 1, the IRS 1 and IRS 2, and the IRS 2 and user are given by $d_\mathrm{BI} = \sqrt{x_\mathrm{BI}^2 + h_\mathrm{s}^2}$, $d_\mathrm{IU} = \sqrt{x_\mathrm{IU}^2 + h_\mathrm{s}^2}$, $d_\mathrm{B} = \sqrt{x_\mathrm{B}^2 + h_\mathrm{d}^2}$, $d_\mathrm{I} = L-x_\mathrm{B}-x_\mathrm{U}$, and $d_\mathrm{U} = \sqrt{x_\mathrm{U}^2 + h_\mathrm{d}^2}$, respectively. 

Let $\mathcal{N}_\mathrm{p}$ and $\mathcal{N}_\mathrm{a}$ denote the sets of all elements on the PIRS and AIRS, respectively. The reflection matrices of PIRS/passive IRS sub-surface and AIRS/active IRS sub-surface are denoted by ${\mathbf{\Psi }}_{{\mathrm{p}}} = \operatorname{diag} ( e^{j\phi_{{\mathrm{p}},1}}, \ldots, e^{j\phi_{{\mathrm{p}},N_\mathrm{p}}} )$ and ${\mathbf{\Psi }}_{\mathrm{a}} = {\mathbf{A}}_{{\mathrm{a}}} \mathbf{\Phi}_{\mathrm{a}}$, respectively, where ${\mathbf{A}}_{{\mathrm{a}}} = \operatorname{diag} (\alpha_{\mathrm{a},1}, \ldots, \alpha_{\mathrm{a},N_\mathrm{a}})$ and ${\mathbf{\Phi }}_{{\mathrm{a}}} = \operatorname{diag} ( e^{j\phi_{{\mathrm{a}},1}}, \ldots, e^{j\phi_{{\mathrm{a}},N_\mathrm{a}}} )$ respectively denote the amplification matrix and phase-shift matrix. $\phi_{{\mathrm{p}},n}, \forall n \in \mathcal{N}_\mathrm{p} (\phi_{{\mathrm{a}},n}, \forall n \in \mathcal{N}_\mathrm{a})$ and $\alpha_{\mathrm{a},n}, \forall n \in \mathcal{N}_\mathrm{a}$ represent the passive (active) phase shift and the active amplification factor at element $n$, respectively. We define the receive response vector as $\mathbf{a}_\mathrm{r} (\theta^\mathrm{r}_{i,k},\vartheta ^\mathrm{r}_{i,k},N_\mathrm{p}) = \boldsymbol{w}(\frac{2\Delta_\mathrm{d}}{\lambda} \sin (\theta^\mathrm{r}_{i,k}) \sin (\vartheta^\mathrm{r}_{i,k}), N_{k,x}) \otimes \boldsymbol{w}(\frac{2\Delta_\mathrm{d}}{\lambda} \cos (\vartheta^\mathrm{r}_{i,k}), N_{k,y})$, where $\lambda$ denotes the signal wavelength, $\Delta_\mathrm{d}$ represents the reflecting element space, $\theta^\mathrm{r}_{i,k}(\vartheta ^\mathrm{r}_{i,k})$ is the azimuth (elevation) angle-of-arrival at node $k$ from node $i$, $N_{k,x}(N_{k,y})$ denotes the number of horizon (vertical) elements at node $k$ with $N_{k} = N_{k,x} \times N_{k,y}$, and $\boldsymbol{w}(\theta,N') = [1, \ldots, e^{-j \pi(N'-1)\theta}]^{T}$ denotes the steering vector function. The transmit steering vector $\mathbf{a}_\mathrm{t}$ can be defined similarly as $\mathbf{a}_\mathrm{r}$. Next, the system models for the considered schemes are described in detail.

\subsection{BHU Scheme}
For BHU scheme, the channel from BS to the passive IRS sub-surface and that from the passive IRS sub-surface to the user are denoted by $\mathbf{h}^\mathrm{p}_\mathrm{BI} = \beta_\mathrm{BI} \mathbf{a}_\mathrm{r} (\theta^\mathrm{r}_{\mathrm{BS},\mathrm{H}},\vartheta ^\mathrm{r}_{\mathrm{BS},\mathrm{H}},N_\mathrm{p})$ and $\mathbf{h}^\mathrm{p}_\mathrm{IU} = \beta_\mathrm{IU} \mathbf{a}_\mathrm{t} (\theta^\mathrm{t}_{\mathrm{H},\mathrm{U}},\vartheta^\mathrm{t}_{\mathrm{H},\mathrm{U}},N_\mathrm{p})$, respectively, where $\beta_\mathrm{BI}$ and $\beta_\mathrm{IU}$ denote the corresponding complex-valued path gains. The channel from BS to active IRS sub-surface $\mathbf{h}^\mathrm{a}_\mathrm{BI}$ and the channel from active IRS sub-surface to the user $\mathbf{h}^\mathrm{a}_\mathrm{IU}$ can be modeled similarly to $\mathbf{h}^\mathrm{p}_\mathrm{BI}$ and $\mathbf{h}^\mathrm{p}_\mathrm{IU}$. Let $P_\mathrm{I}$ denote the amplification power budget of the active elements and then we have 
\begin{align}
	{P_{\mathrm{B}}}{\left\| {{{\mathbf{\Psi }}_{{\mathrm{a}}}}{\mathbf{h}}_{{\mathrm{BI}}}^{{\mathrm{a}}}} \right\|^2} + \sigma _{\mathrm{r}}^2{\left\| {{{\mathbf{\Psi }}_{{\mathrm{a}}}}{{\mathbf{I}}_{{N_{{\mathrm{a}}}}}}} \right\|^2} \le {P_{\mathrm{I}}},
\end{align}
where $\sigma_{\mathrm{r}}^2$ denotes the amplification noise power.
Though the single-reflection link, the signal received at the user is given by ${y_{\mathrm{0}}} = {\left( {{\mathbf{h}}_{{\mathrm{IU}}}^{{\mathrm{a}}}} \right)^H}{{\mathbf{\Psi}}_{{\mathrm{a}}}}{\mathbf{h}}_{{\mathrm{BI}}}^{{\mathrm{a}}}s + {\left( {{\mathbf{h}}_{{\mathrm{IU}}}^{{\mathrm{p}}}} \right)^H}{{\mathbf{\Psi }}_{{\mathrm{p}}}}{\mathbf{h}}_{{\mathrm{BI}}}^{{\mathrm{p}}}s + {\left( {{\mathbf{h}}_{{\mathrm{IU}}}^{{\mathrm{a}}}} \right)^H}{{\mathbf{\Psi }}_{{\mathrm{a}}}}{{\mathbf{n}}_{\mathrm{r}}} + {n_0}$, where $s \in \mathbb{C}$ represents the transmitted signal satisfying $\mathbb{E} \{ {{{\left| s \right|}^2}} \} = P_\mathrm{B}$ with the BS transmit power $P_\mathrm{B}$, ${\mathbf{n}}_{\mathrm{r}} \sim \mathcal{CN} (\mathbf{0}_{N_\mathrm{a}}, \sigma_{\mathrm{r}}^2 \mathbf{I}_{N_\mathrm{a}})$ denotes the amplification noise introduced by active elements, and $n_0 \sim \mathcal{CN}(0, \sigma_0^2)$ denotes the additive white Gaussian noise (AWGN) at the user. Accordingly, the SNR at the user is given by
\begin{align}
	\label{SNR_H}
	{\gamma _{\mathrm{0}}} = \frac{{{P_{\mathrm{B}}}{{\left| {{{\left( {{\mathbf{h}}_{{\mathrm{IU}}}^{{\mathrm{a}}}} \right)^H}}{{\mathbf{\Psi }}_{{\mathrm{a}}}}{\mathbf{h}}_{{\mathrm{BI}}}^{{\mathrm{a}}} + {{\left( {{\mathbf{h}}_{{\mathrm{IU}}}^{{\mathrm{p}}}} \right)^H}}{{\mathbf{\Psi }}_{{\mathrm{p}}}}{\mathbf{h}}_{{\mathrm{BI}}}^{{\mathrm{p}}}} \right|}^2}}}{{\sigma _{\mathrm{r}}^2{{\left\| {{{\left( {{\mathbf{h}}_{{\mathrm{IU}}}^{{\mathrm{a}}}} \right)^H}}{{\mathbf{\Psi }}_{{\mathrm{a}}}}} \right\|}^2} + \sigma _0^2}}, 
\end{align}
and the achievable rate for the BHU scheme is expressed as $R_0 = {{\log _2}\left( {1 + {\gamma _{\mathrm{0}}}} \right)}$. 
The phase of the HIRS should be designed to align in the BS-HIRS-user channels, i.e. ${{\left( {{\mathbf{h}}_{{\mathrm{IU}}}^{{\mathrm{a}}}} \right)^H}}{{\mathbf{\Psi }}_{{\mathrm{a}}}}{\mathbf{h}}_{{\mathrm{BI}}}^{{\mathrm{a}}}$ and ${{\left( {{\mathbf{h}}_{{\mathrm{IU}}}^{{\mathrm{p}}}} \right)^H}}{{\mathbf{\Psi }}_{{\mathrm{p}}}}{\mathbf{h}}_{{\mathrm{BI}}}^{{\mathrm{p}}}$. According to \cite{HIRS_SE}, the optimal HIRS beamforming can be expressed as $\phi_{{\mathrm{p}},n} = \arg \left( {{{[ {{\mathbf{h}}_{{\mathrm{IU}}}^{{\mathrm{p}}}} ]_n}}} \right) - \arg \left( {{{[ {{\mathbf{h}}_{{\mathrm{BI}}}^{{\mathrm{p}}}} ]_n}}} \right)$, $\phi_{{\mathrm{a}},n} = \arg \left( {{{ [ {{\mathbf{h}}_{{\mathrm{IU}}}^{{\mathrm{a}}}} ]_n}}} \right) - \arg \left( {{{ [ {{\mathbf{h}}_{{\mathrm{BI}}}^{{\mathrm{a}}}} ]_n}}} \right)$, and $\alpha_{\mathrm{a},n} = {\alpha _{\mathrm{0}}} \triangleq \sqrt {{{{P_{\mathrm{I}}}}}/({{{N_{{\mathrm{a}}}}\left( {{P_{\mathrm{B}}}\beta _{{\mathrm{BI}}}^2 + \sigma _{\mathrm{r}}^2} \right)}})}$, \footnote{To facilitate comparison, we consider the case where  $P_\mathrm{I}$ satisfies the favorable amplification power condition similar to \cite{HIRS_SE}, i.e., $P_\mathrm{I} \le \min \{ (N-1){{\left( {{P_{\mathrm{B}}}\beta _{{\mathrm{BI}}}^2 + \sigma _{\mathrm{r}}^2} \right)}}, (N-1) {{ ( {{P_{\mathrm{B}}}\beta _{\mathrm{B}}^2 + \sigma _{\mathrm{r}}^2} )}}, (N-1) ( {{P_{\mathrm{B}}}\beta _{\mathrm{I}}^2\beta _{\mathrm{B}}^2 + \sigma _{\mathrm{r}}^2} ) \}$, ensuring that $\alpha_{\mathrm{a},n} \le \alpha_\mathrm{max}$ always holds, which is constrained by the limited load of each active element \cite{a_max}.} which leads to 
\begin{align}
	\label{SNR_H_cf}
	{\gamma _{\mathrm{0}}} = \frac{{{P_{\mathrm{B}}}\beta _{{\mathrm{BI}}}^2\beta _{{\mathrm{IU}}}^2 {{{\left( {\sqrt {{N_{{\mathrm{a}}}}{P_{\mathrm{I}}}}  + \sqrt {N_{{\mathrm{p}}}^2\left( {{P_{\mathrm{B}}}\beta _{{\mathrm{BI}}}^2 + \sigma _{\mathrm{r}}^2} \right)} } \right)^2}}}}}{{\sigma _{\mathrm{r}}^2{P_{\mathrm{I}}}\beta _{{\mathrm{IU}}}^2 + \sigma _0^2{P_{\mathrm{B}}}\beta _{{\mathrm{BI}}}^2 + \sigma _0^2\sigma _{\mathrm{r}}^2}}.
\end{align}
We aim to maximize the achievable rate by jointly optimizing the HIRS elements allocation $\mathbf{N} \buildrel \Delta \over = \left\{N_\mathrm{p}, N_\mathrm{a}\right\}$ and placement $\mathbf{X}_\mathrm{s} \buildrel \Delta \over = \left\{x_\mathrm{BI}, x_\mathrm{IU}\right\}$. The optimization problem is written as
\begin{subequations}
	\label{Probem_H}
	\begin{align}
		\mathop {\max }\limits_{{\mathbf{N}, \mathbf{X}_\mathrm{s}}} \;\;\; & R_0  \\
		{\mathrm{s.t.}} \;\;\;\;
		& N_\mathrm{p} + N_\mathrm{a} = N, \label{con_Q} \\
		& N_\mathrm{p} \in \mathbb{N}^+, N_\mathrm{a} \in \mathbb{N}^+, \label{con_N} \\
		& x_\mathrm{BI} + x_\mathrm{IU} = L, \label{x_h_1} \\
		& x_\mathrm{BI} \ge 0, x_\mathrm{IU} \ge 0. \label{x_h_2}
	\end{align}
\end{subequations}

\subsection{BAPU Scheme}
For BAPU scheme, the channels from BS to the AIRS, from AIRS to PIRS, and from PIRS to user are denoted by $\mathbf{h}_\mathrm{BA} = \beta_\mathrm{B} \mathbf{a}_\mathrm{r} (\theta^\mathrm{r}_{\mathrm{BS},\mathrm{A}},\vartheta ^\mathrm{r}_{\mathrm{BS},\mathrm{A}},N_\mathrm{a})$, ${{\mathbf{H}}_{{\mathrm{AP}}}} = {\beta _{\mathrm{I}}}{{\mathbf{a}}_{\mathrm{r}}}\left( {\theta _{{\mathrm{BS,A}}}^{\mathrm{r}},\vartheta _{{\mathrm{BS,A}}}^{\mathrm{r}},{N_{\mathrm{a}}}} \right){\mathbf{a}}_{\mathrm{t}}^H\left( {\theta _{{\mathrm{P,U}}}^{\mathrm{r}},\vartheta _{{\mathrm{P,U}}}^{\mathrm{r}},{N_{\mathrm{p}}}} \right)$, and $\mathbf{h}_\mathrm{PU} = \beta_\mathrm{U} \mathbf{a}_\mathrm{t} (\theta^\mathrm{t}_{\mathrm{P},\mathrm{U}},\vartheta^\mathrm{t}_{\mathrm{P},\mathrm{U}},N_\mathrm{p})$, respectively, where $\beta_\mathrm{B}^2$, $\beta_\mathrm{I}^2$, and $\beta_\mathrm{U}^2$ denote the corresponding path loss. The amplification power constraint is given by
\begin{align}
	{P_{\mathrm{B}}}{\left\| {{{\mathbf{\Psi }}_{{\mathrm{a}}}}{{\mathbf{h}}_{\mathrm{BA}}}} \right\|^2} + \sigma _{\mathrm{r}}^2{\left\| {{{\mathbf{\Psi }}_{{\mathrm{a}}}}{{\mathbf{I}}_{{N_{{\mathrm{a}}}}}}} \right\|^2} \le {P_{\mathrm{I}}}.
\end{align}
Through the BS→AIPS→PIRS→user link, the received signal at the user is expressed as ${y_1} = {\mathbf{h}}_{\mathrm{PU}}^H{{\mathbf{\Psi}}_{{\mathrm{p}}}}{{\mathbf{H}}_{{\mathrm{AP}}}}{{\mathbf{\Psi}}_{{\mathrm{a}}}}{{\mathbf{h}}_{\mathrm{BA}}}s + {\mathbf{h}}_{\mathrm{PU}}^H{{\mathbf{\Psi}}_{{\mathrm{p}}}}{{\mathbf{H}}_{{\mathrm{AP}}}}{{\mathbf{\Psi }}_{{\mathrm{a}}}}{{\mathbf{n}}_{\mathrm{r}}} + {n_0}$. Accordingly, the SNR of the user is given by
\begin{align}
	\label{SNR_AP}
	{\gamma _1} = \frac{{{P_{\mathrm{B}}}{{\left| {{\mathbf{h}}_{\mathrm{PU}}^H{{\mathbf{\Psi }}_{{\mathrm{p}}}}{{\mathbf{H}}_{{\mathrm{AP}}}}{{\mathbf{\Psi }}_{{\mathrm{a}}}}{{\mathbf{h}}_{\mathrm{BA}}}} \right|}^2}}}{{\sigma _{\mathrm{r}}^2{{\left\| {{\mathbf{h}}_{\mathrm{PU}}^H{{\mathbf{\Psi }}_{{\mathrm{p}}}}{{\mathbf{H}}_{{\mathrm{AP}}}}{{\mathbf{\Psi }}_{{\mathrm{a}}}}} \right\|}^2} + \sigma _0^2}},
\end{align}
and the achievable rate for the BAPU scheme is expressed as $R_1 = {{\log _2}\left( {1 + {\gamma _1}} \right)}$. The phases of the AIRS and PIRS is optimized to align in the BS-AIRS-PIRS-user channels, i.e. ${{\mathbf{h}}_{\mathrm{PU}}^H{{\mathbf{\Psi }}_{{\mathrm{p}}}}{{\mathbf{H}}_{{\mathrm{AP}}}}{{\mathbf{\Psi }}_{{\mathrm{a}}}}{{\mathbf{h}}_{\mathrm{BA}}}}$. According to \cite{AP_x}, the optimal IRS beamforming design for BAPU scheme is expressed as $\phi_{{\mathrm{a}},n} = \arg ( {{{[ {{{\mathbf{a}}_{\mathrm{t}}}( {\theta _{\mathrm{P,U}}^{\mathrm{r}},\vartheta _{\mathrm{P,U}}^{\mathrm{r}},{N_{{\mathrm{a}}}}} )} ]_n}}} ) - \arg ( {{{[ {{{\mathbf{h}}_{\mathrm{BA}}}} ] _n}}} )$, $\phi_{{\mathrm{p}},n} = \arg ( {{{ [ {{{\mathbf{h}}_{\mathrm{PU}}}} ] _n}}} ) - \arg ( {{{[ {{{\mathbf{a}}_{\mathrm{r}}} ( {\theta _{\mathrm{BS,A}}^{\mathrm{r}},\vartheta _{\mathrm{BS,A}}^{\mathrm{r}},{N_{{\mathrm{p}}}}} )} ]_n}}} )$, and $\alpha_{\mathrm{a},n} = {\alpha _1} \triangleq \sqrt {{{{P_{\mathrm{I}}}}}/({{{N_{{\mathrm{a}}}}\left( {{P_{\mathrm{B}}}\beta _{\mathrm{B}}^2 + \sigma _{\mathrm{r}}^2} \right)}})}$, which leads to
\begin{align}
	\label{SNR_AP_cf}
	{\gamma _1} = \frac{{\beta _{\mathrm{B}}^2\beta _{\mathrm{I}}^2\beta _{\mathrm{U}}^2{P_{\mathrm{I}}}{P_{\mathrm{B}}}{N_{{\mathrm{a}}}}N_{{\mathrm{p}}}^2}}{{\sigma _{\mathrm{r}}^2\beta _{\mathrm{U}}^2\beta _{\mathrm{I}}^2{P_{\mathrm{I}}}{N_{{\mathrm{p}}}^2} + \sigma _0^2{P_{\mathrm{B}}}\beta _{\mathrm{B}}^2 + \sigma _0^2\sigma _{\mathrm{r}}^2}}.
\end{align}
Similar to the BHU scheme, the optimization problem for the BAPU scheme with the placement $\mathbf{X}_\mathrm{d} \buildrel \Delta \over = \left\{x_\mathrm{B}, x_\mathrm{I}, x_\mathrm{U}\right\}$ is formulated as
\begin{subequations}
	\label{Probem_AP}
	\begin{align}
		\mathop {\max }\limits_{{\mathbf{N}, \mathbf{X}_\mathrm{d}}} \;\;\; &  R_1 \\
		{\mathrm{s.t.}} \;\;\;\;
		& x_\mathrm{B} + x_\mathrm{I} + x_\mathrm{U} = L, \label{x_1} \\
		& x_\mathrm{B} \ge 0, x_\mathrm{I} \ge 0, x_\mathrm{U} \ge 0, \label{x_2} \\
		&\eqref{con_Q}, \eqref{con_N} \nonumber.
	\end{align}
\end{subequations}
\subsection{BPAU Scheme}
For BPAU scheme, the channels between different nodes can be modeled similarly to those for BAPU scheme, which are omitted for brevity. Since the AIRS and PIRS are placed in reverse order, the amplification power constraint is given by
\begin{align}
	{P_{\mathrm{B}}}{\left\| {{{\mathbf{\Psi }}_{{\mathrm{a}}}}{{\mathbf{H}}_{{\mathrm{PA}}}}{{\mathbf{\Psi }}_{{\mathrm{p}}}}{{\mathbf{h}}_{\mathrm{BP}}}} \right\|^2} + \sigma _{\mathrm{r}}^2{\left\| {{{\mathbf{\Psi }}_{{\mathrm{a}}}}{{\mathbf{I}}_{{N_{{\mathrm{a}}}}}}} \right\|^2} \le {P_{\mathrm{I}}}.
\end{align}
Through the BS→PIPS→AIRS→user link, the received signal at the user is expressed as ${y_2} = {\mathbf{h}}_{\mathrm{AU}}^H{{\mathbf{\Psi }}_{{\mathrm{a}}}}{{\mathbf{H}}_{{\mathrm{PA}}}}{{\mathbf{\Psi }}_{{\mathrm{p}}}}{{\mathbf{h}}_{\mathrm{BP}}}s + {\mathbf{h}}_{\mathrm{AU}}^H{{\mathbf{\Psi }}_{{\mathrm{a}}}}{{\mathbf{n}}_{\mathrm{r}}} + {n_0}$. Accordingly, the SNR of the user is given by
\begin{align}
	\label{SNR_PA}
	{\gamma _2} = \frac{{{P_{\mathrm{B}}}{{\left| {{\mathbf{h}}_{\mathrm{AU}}^H{{\mathbf{\Psi }}_{{\mathrm{a}}}}{{\mathbf{H}}_{{\mathrm{PA}}}}{{\mathbf{\Psi }}_{{\mathrm{p}}}}{{\mathbf{h}}_{\mathrm{BP}}}} \right|}^2}}}{{\sigma _{\mathrm{r}}^2{{\left\| {{\mathbf{h}}_{\mathrm{AU}}^H{{\mathbf{\Psi }}_{{\mathrm{a}}}}} \right\|}^2} + \sigma _0^2}},
\end{align}
and the achievable rate for the BPAU scheme is expressed as $R_2 = {{\log _2}\left( {1 + {\gamma _2}} \right)}$. The phases of the AIRS and PIRS is optimized to align in the BS-AIRS-PIRS-user channels, i.e. ${{\mathbf{h}}_{\mathrm{AU}}^H{{\mathbf{\Psi }}_{{\mathrm{a}}}}{{\mathbf{H}}_{{\mathrm{PA}}}}{{\mathbf{\Psi }}_{{\mathrm{p}}}}{{\mathbf{h}}_{\mathrm{BP}}}}$. Similarly, the optimal IRS beamforming design for BPAU scheme is expressed as $\phi_{{\mathrm{p}},n} = \arg ( {{{ [ {{{\mathbf{a}}_{\mathrm{t}}} ( {\theta _{\mathrm{BS,P}}^{\mathrm{r}},\vartheta _{\mathrm{BS,P}}^{\mathrm{r}},{N_{{\mathrm{p}}}}} )} ] _n}}} ) - \arg ( {{{ [ {{{\mathbf{h}}_{\mathrm{BP}}}} ] _n}}} )$, $\phi_{{\mathrm{a}},n} = \arg ( {{{ [ {{{\mathbf{h}}_{\mathrm{AU}}}} ] _n}}} ) - \arg ( {{{ [ {{{\mathbf{a}}_{\mathrm{r}}}\left( {\theta _{\mathrm{A,U}}^{\mathrm{r}},\vartheta _{\mathrm{A,U}}^{\mathrm{r}},{N_{{\mathrm{a}}}}} \right)} ] _n}}} )$, and $\alpha_{\mathrm{a},n} = {\alpha _2} \triangleq \sqrt {{{{P_{\mathrm{I}}}}}/({{{N_{{\mathrm{a}}}}\left( {{P_{\mathrm{B}}}\beta _{\mathrm{I}}^2\beta _{\mathrm{B}}^2{N_{{\mathrm{p}}}^2} + \sigma _{\mathrm{r}}^2} \right)}})}$, which leads to
\begin{align}
	\label{SNR_PA_cf}
	{\gamma _2} = \frac{{\beta _{\mathrm{B}}^2\beta _{\mathrm{I}}^2\beta _{\mathrm{U}}^2{P_{\mathrm{I}}}{P_{\mathrm{B}}}{N_{{\mathrm{a}}}}N_{{\mathrm{p}}}^2}}{{\sigma _0^2{P_{\mathrm{B}}}\beta _{\mathrm{I}}^2\beta _{\mathrm{B}}^2{N_{{\mathrm{p}}}^2} + \sigma _{\mathrm{r}}^2\beta _{\mathrm{U}}^2{P_{\mathrm{I}}} + \sigma _0^2\sigma _{\mathrm{r}}^2}}.
\end{align}
Accordingly, the optimization problem for the BPAU scheme is written as
\begin{subequations}
	\label{Probem_PA}
	\begin{align}
		\mathop {\max }\limits_{{\mathbf{N}, \mathbf{X}_\mathrm{d}}} \;\;\; &  R_2 \\
		{\mathrm{s.t.}} \;\;\;\;
		&\eqref{con_Q}, \eqref{con_N}, \eqref{x_1}, \eqref{x_2} \nonumber.
	\end{align}
\end{subequations}

Note that problems \eqref{Probem_H}, \eqref{Probem_AP}, and \eqref{Probem_PA} are all non-convex and difficult to solve optimally due to the integer variables coupled in the non-convex objective function. To overcome these challenges, we solve them by decomposing the problem into two sub-problems in the next section.

\begin{Remark}	In mobile scenarios where the user moves in the hotspot area, the user trajectory can be predicted via the standard Kalman filter method to forecast the corresponding channel. In particular, under the BAPU and BPAU schemes, the BS sends the transmit signal to IRS 1, which then reflects the signal towards IRS 2, similar to the considered static scenario. On the other hand, regarding the channel variation from IRS 2 to user caused by user mobility, the beamforming of HIRS and IRS 2 can be dynamically adjusted to improve the achievable rate. To our best understanding, the proposed deployment strategies are still applicable in this case. To be aggressive, movable IRS can also be deployed to address the user mobility issue, where the IRS deployment can follow similar scanning-based optimization methods as in \cite{R2}. Considering the non-trivial extension of this method, we have to leave it for our future work.
\end{Remark}

\section{IRS Elements Allocation and Placement Optimization}
\label{EAP_opt}
In this section, we focus on the IRS elements allocation and placement design to maximize the achievable rate. Given the optimal IRS beamforming, the IRS elements allocation and placement are jointly optimized. To draw essential insights, we first derive the optimized solution in closed form and then provide a theoretical performance comparison of the three IRS deployment schemes in terms of the achievable rate.

\subsection{Active/Passive Elements Allocation Optimization}
For any given IRS placement, we first optimize the IRS elements allocation. Based on \eqref{SNR_H_cf}, \eqref{SNR_AP_cf}, and \eqref{SNR_PA_cf}, for the case with BHU scheme ($i = \mathrm{0}$), BAPU scheme ($i = \mathrm{1}$), and BPAU scheme ($i = \mathrm{2}$), the optimization problem is reformulated as
\begin{subequations}
	\label{Problem_elements}
	\begin{align}
		\mathop {\max }\limits_{{N_\mathrm{p}},{N_\mathrm{a}}} \;\; &\gamma _i\\
		{\mathrm{s.t.}} \;\;\;		
		& N_\mathrm{p} + N_\mathrm{a} = N, \label{con_Q1} \\
		& N_\mathrm{p} \in \mathbb{N}^+, N_\mathrm{a} \in \mathbb{N}^+.
	\end{align}
\end{subequations}
Problem \eqref{Problem_elements} is still intractable due to the non-concave objective function and the discrete variables in constraint \eqref{con_Q1}, making it challenging to obtain the optimal solution in closed form. Although the optimal solution to problem \eqref{Problem_elements} can be obtained through numerical methods like one-dimensional search over ${N_\mathrm{p}} = \left\{ 1, \ldots, N - 1 \right\}$ with ${N_\mathrm{a}} = N - {N_\mathrm{p}}$, it fails to offer valuable insight into the optimal IRS elements allocation. 

To address these issues, we first relax the integer values $N_\mathrm{p}$ and $N_\mathrm{a}$ into their continuous counterparts, denoted by $\tilde N_\mathrm{p}$ and $\tilde N_\mathrm{a}$. Then, the optimized solution to problem \eqref{Problem_elements} can be reconstructed by the integer rounding technique. As such, we derive the following propositions to obtain optimized active/passive elements allocation under the three schemes and then characterize the impacts of key system parameters based on the closed-from expressions.

\subsubsection{BHU Scheme} We derive the closed-form expression of the optimal solution to problem \eqref{Problem_elements} when $i=0$, i.e., BHU scheme, as detailed in the following.
\begin{Proposition}
	\label{pro_N_H}
	For BHU scheme, the optimal number of active and passive elements at the HIRS are given by	
	\begin{align}
		\label{N_h}
		\left\{ {\begin{array}{*{20}{c}}
				{{\tilde N_{{\mathrm{a}}}^* = N-1,\tilde N_{{\mathrm{p}}}^* = 1},}&{{{N \le \frac{{{P_{\mathrm{I}}}}}{{4 {{P_{\mathrm{B}}}\beta _{{\mathrm{BI}}}^2 + 4 \sigma _{\mathrm{r}}^2} }}}},}\\
				\begin{array}{l}
					{\tilde N_{{\mathrm{a}}}^* = \frac{{{P_{\mathrm{I}}}}}{{4 {{P_{\mathrm{B}}}\beta _{{\mathrm{BI}}}^2 + 4\sigma _{\mathrm{r}}^2}}},}\\
					\tilde N_{{\mathrm{p}}}^* = N - \frac{{{P_{\mathrm{I}}}}}{{4 {{P_{\mathrm{B}}}\beta _{{\mathrm{BI}}}^2 + 4\sigma _{\mathrm{r}}^2}}},
				\end{array}&{\text{Otherwise.}}
		\end{array}} \right.
	\end{align}
\end{Proposition}

{\it{Proof:}} 
Please refer to Appendix A. 
~$\hfill\blacksquare$

From Proposition \ref{pro_N_H}, one can observe that more active elements should be deployed at the HIRS when the total number of IRS elements $N$ is small, benefiting from its amplification gain. Otherwise, $\tilde N_\mathrm{a}^*$ is independent of the total number of IRS elements $N$ due to the limited amplification power budget, which becomes the performance bottleneck for active elements. In this case, $\tilde N_\mathrm{a}^*$ is determined by other parameters, e.g., ${P_{\mathrm{I}}}$ and $d_\mathrm{BI}$. Specifically, $\tilde N_\mathrm{a}^*$ monotonically increases with the amplification power ${P_{\mathrm{I}}}$ and the BS-IRS distance $d_\mathrm{BI}$. The reasons are given as follows. First, a higher amplification power budget ${P_{\mathrm{I}}}$ enables a larger number of active elements to operate with sufficiently large amplification factors. Second, since a longer BS-IRS distance leads to severe path loss, more active elements should be placed to provide a higher amplification gain, thereby compensating for the attenuation. By substituting \eqref{N_h} into \eqref{SNR_H_cf}, the SNR with optimal active/passive elements allocation is given by
\begin{align}
	\label{SNR_H_N}
	{{\gamma _\mathrm{0,N}^*}} = 
	\left\{ {\begin{array}{*{20}{l}}
			{{{c_1}{{( {{c_2}\sqrt {N - 1}  + 1} )^2}}},}\\
			{{c_1}(c_2^2/4+N)^2,}
		\end{array}\begin{array}{*{20}{l}}
			{{N \le \frac{{{P_{\mathrm{I}}}}}{{4 {{P_{\mathrm{B}}}\beta _{{\mathrm{BI}}}^2 + 4 \sigma _{\mathrm{r}}^2} }}},}\\
			\text{Otherwise,}
	\end{array}} \right. 
\end{align}
with $c_1 = \frac{{{P_{\mathrm{B}}}\beta _{{\mathrm{BI}}}^2\beta _{{\mathrm{IU}}}^2\left( {{P_{\mathrm{B}}}\beta _{{\mathrm{BI}}}^2 + \sigma _{\mathrm{r}}^2} \right)}}{{ {\sigma _{\mathrm{r}}^2{P_{\mathrm{I}}}\beta _{{\mathrm{IU}}}^2 + \sigma _0^2{P_{\mathrm{B}}}\beta _{{\mathrm{BI}}}^2 + \sigma _0^2\sigma _{\mathrm{r}}^2} }}$ and $c_2 = \sqrt {\frac{{{P_{\mathrm{I}}}}}{{ {{P_{\mathrm{B}}}\beta _{{\mathrm{BI}}}^2 + \sigma _{\mathrm{r}}^2} }}}$. From \eqref{SNR_H_N}, it is observed that the SNR first linearly increases with $N$ and then increases with $N$ as $\mathcal{O}\left(N^2\right)$ in the large-$N$ regime.

\subsubsection{BAPU Scheme} We derive the optimal solution to problem \eqref{Problem_elements} when $i=1$, i.e., BAPU scheme, which is provided as follows.
\begin{Proposition}
	\label{pro_N_AP}
	For BAPU scheme, the optimal number of passive elements is given by ${{\tilde N}_{{\mathrm{p}}}^*} = x_1^\mathrm{rt}$, where $x_1^\mathrm{rt}$ is the unique solution to the equation
	\begin{align}
		\label{g1(x)}
		g_1\left(x\right) \buildrel \Delta \over = { - {c_3}x^3 - 3{c_4}x + 2N{c_4}} = 0,
	\end{align}
	with ${c_3} = \sigma _{\mathrm{r}}^2\beta _{\mathrm{U}}^2\beta _{\mathrm{I}}^2{P_{\mathrm{I}}}$ and ${c_4} = {\sigma _0^2{P_{\mathrm{B}}}\beta _{\mathrm{B}}^2 + \sigma _0^2\sigma _{\mathrm{r}}^2}$. 
\end{Proposition}

{\it{Proof:}}
Please refer to Appendix B.
~$\hfill\blacksquare$

Since \eqref{g1(x)} is a cubic equation, $x_1^\mathrm{rt}$ can be calculated by the Cardano’s formula. From Proposition \ref{pro_N_AP}, it can be readily verified that $\tilde N_\mathrm{a}^*$ monotonically increases with the amplification power budget ${P_{\mathrm{I}}}$ by ensuring that the equation \eqref{g1(x)} holds. The result is the same as that of the BHU scheme, thus the detailed explanation is omitted for brevity. Moreover, more passive elements should be deployed at the PIRS to provide a higher beamforming gain for mitigating the severe two-hop path-loss as $\beta _{\mathrm{U}}^2\beta _{\mathrm{I}}^2$ increases. Assuming that ${c_3} \tilde{N}_{{\mathrm{p}}}^2 \ll {c_4}$, $\gamma _1$ can be rewritten as $\gamma _1 = {c_6}(N-{{\tilde N}_{{\mathrm{p}}}}){{\tilde N}_{{\mathrm{p}}}}^2/{c_4}$, with $c_6 = {\beta _{\mathrm{B}}^2\beta _{\mathrm{I}}^2\beta _{\mathrm{U}}^2{P_{\mathrm{I}}}{P_{\mathrm{B}}}}$. Therefore, the solution to problem \eqref{Problem_elements} is obtained by setting the first-order partial derivative of $\gamma _{\mathrm{1}}$ w.r.t. $\tilde N_\mathrm{p}$ to zero, which is given by ${{\tilde N}_{{\mathrm{a}}}^*} = N/3, {{\tilde N}_{{\mathrm{p}}}^*} = 2N/3$. As such, the approximated SNR ${{\gamma _\mathrm{1,N}^*}} = \frac{4{c_6}}{27{c_4}}N^3$ increases with $N$ as $\mathcal{O}\left(N^3\right)$, which exceeds the scaling for the BHU scheme in the large-$N$ regime. It is not difficult to verified that ${{\gamma _\mathrm{1,N}^*}}$ linearly increases with ${P_{\mathrm{I}}}$ and also increases with ${P_{\mathrm{B}}}$, but the rate of the latter is slower than that of the former. The results unveil that increasing the amplification power is more effective for the BAPU scheme to improve the performance in terms of achievable rate.

\subsubsection{BPAU Scheme} We derive the optimal solution to problem \eqref{Problem_elements} when $i=2$, i.e., BPAU scheme, as detailed in the following proposition.
\begin{Proposition}
	\label{pro_N_PA}
	For BPAU scheme, the optimal number of passive elements is given by ${{\tilde N}_{{\mathrm{p}}}^*} = x_2^\mathrm{rt}$, where $x_2^\mathrm{rt}$ is the unique solution to the equation
	\begin{align}
		\label{g2(x)}
		g_2\left(x\right) \buildrel \Delta \over = { -{c_7}x^3 - 3{c_8}x + 2N{c_8}} = 0,
	\end{align}
	with ${c_7} = \sigma _0^2{P_{\mathrm{B}}}\beta _{\mathrm{I}}^2\beta _{\mathrm{B}}^2$ and ${c_8} = {\sigma _{\mathrm{r}}^2\beta _{\mathrm{U}}^2{P_{\mathrm{I}}} + \sigma _0^2\sigma _{\mathrm{r}}^2}$. 
\end{Proposition}

{\it{Proof:}} The proof is similar to that in Appendix B and is thus omitted for brevity. ~$\hfill\blacksquare$

Since \eqref{g2(x)} is a cubic equation, $x_2^\mathrm{rt}$ can be calculated by the Cardano’s formula. From Proposition \ref{pro_N_PA}, it is observed that more active elements should be deployed at the AIRS as ${P_{\mathrm{B}}}$ or $\beta _{\mathrm{I}}^2\beta _{\mathrm{B}}^2$ increases. This is because the amplification gain can be maintained even though the amplification factor $\alpha_2$ respectively decreases with ${P_{\mathrm{B}}}$ and $\beta _{\mathrm{I}}^2\beta _{\mathrm{B}}^2$. Assuming that ${c_7} \tilde{N}_{{\mathrm{p}}}^2 \ll {c_8}$, $\gamma _1$ can be rewritten as $\gamma _2 = {c_6} (N-{{\tilde N}_{{\mathrm{p}}}} ){{\tilde N}_{{\mathrm{p}}}}^2/{c_8}$. Therefore, the solution to problem \eqref{Problem_elements} is obtained by setting the first-order partial derivative of $\gamma _{\mathrm{2}}$ w.r.t. $\tilde N_\mathrm{p}$ to zero, which is given by ${{\tilde N}_{{\mathrm{a}}}^*} = N/3, {{\tilde N}_{{\mathrm{p}}}^*} = 2N/3$. For the BPAU scheme, the approximated SNR ${{\gamma _\mathrm{2,N}^*}} = \frac{4{c_6}}{27{c_8}}N^3$ increases with $N$ as $\mathcal{O}\left(N^3\right)$, which is the same as that of the BAPU scheme. Furthermore, it is observed that ${{\gamma _\mathrm{2,N}^*}}$ linearly increases with ${P_{\mathrm{B}}}$ and increases with ${P_{\mathrm{I}}}$ at a slower rate. The result indicates that increasing the transmit power is more beneficial for the BPAU scheme, which is different from the BAPU scheme.

\begin{figure*}[ht]
	\small
	\begin{equation}
		\begin{cases}
			\text{1)} R_{{\mathrm{0}},{\mathrm{N}}}^* = \max \{ {R_{{\mathrm{0}},{\mathrm{N}}}^*,R_{{\mathrm{1}},{\mathrm{N}}}^*,R_{{\mathrm{2}},{\mathrm{N}}}^*} \} \text{ if } d_{{\mathrm{IU}}}^2 < \min  \left\{ {{{27\beta \beta _{{\mathrm{BI}}}^2{P_{\mathrm{B}}}{c_4}{{( {{c_2}\sqrt {N - 1}  + 1} )^2}}} \over {4{c_6}{N^3}\sigma _0^2}} - \beta c_2^2\sigma _{\mathrm{r}}^2, {{27\beta \beta _{{\mathrm{BI}}}^2{P_{\mathrm{B}}}{c_8}{{( {{c_2}\sqrt {N - 1}  + 1})^2}}} \over {4{c_6}{N^3}\sigma _0^2}} - \beta c_2^2\sigma _{\mathrm{r}}^2} \right\}; \\
			
			\text{2)} R_{1,{\mathrm{N}}}^* = \max \{ {R_{{\mathrm{0}},{\mathrm{N}}}^*,R_{{\mathrm{1}},{\mathrm{N}}}^*,R_{{\mathrm{2}},{\mathrm{N}}}^*} \} \text{ if }
			d_{\mathrm{U}}^2 < \min \left\{ {{{\sigma _{\mathrm{r}}^2{P_{\mathrm{I}}}d_{\mathrm{B}}^2} \over {\sigma _0^2{P_{\mathrm{B}}}}},{{4\beta \beta _{\mathrm{I}}^2\beta _{\mathrm{B}}^2{P_{\mathrm{I}}}{P_{\mathrm{B}}}{N^3}} \over {27{c_1}{c_4}{{( {{c_2}\sqrt {N - 1}  + 1} )^2}}}}} \right\}; \\
			
			\text{3)} R_{2,{\mathrm{N}}}^* = \max \{ {R_{{\mathrm{0}},{\mathrm{N}}}^*,R_{{\mathrm{1}},{\mathrm{N}}}^*,R_{{\mathrm{2}},{\mathrm{N}}}^*} \} \text{ if } d_{\mathrm{B}}^2 < \min \left\{ {{{\sigma _0^2{P_{\mathrm{B}}}d_{\mathrm{U}}^2} \over {\sigma _{\mathrm{r}}^2{P_{\mathrm{I}}}}},{{4\beta \beta _{\mathrm{I}}^2\beta _{\mathrm{U}}^2{P_{\mathrm{I}}}{P_{\mathrm{B}}}{N^3}} \over {27{c_1}{c_8}{{\left( {{c_2}\sqrt {N - 1}  + 1} \right)^2}}}}} \right\}.
		\end{cases} \label{con_1}
	\end{equation}
	\vspace{-10pt}
	{\noindent} \rule[-0pt]{18.3cm}{0.05em}
\end{figure*}

\begin{figure*}[ht]
	\small
	\begin{equation}
		\begin{cases}
			\text{1)} R_{{\mathrm{0}},{\mathrm{N}}}^* = \max \{ {R_{{\mathrm{0}},{\mathrm{N}}}^*,R_{{\mathrm{1}},{\mathrm{N}}}^*,R_{{\mathrm{2}},{\mathrm{N}}}^*} \} \text{ if } d_{{\mathrm{IU}}}^2 < \min \left\{ {{{27\beta \beta _{{\mathrm{BI}}}^2{P_{\mathrm{B}}}{c_4}{{( {{c_2}^2/4 + N} )^2}}} \over {4{c_6}{N^3}\sigma _0^2}} - \beta c_2^2\sigma _{\mathrm{r}}^2,{{27\beta \beta _{{\mathrm{BI}}}^2{P_{\mathrm{B}}}{c_8}{{( {{c_2}^2/4 + N} )^2}}} \over {4{c_6}{N^3}\sigma _0^2}} - \beta c_2^2\sigma _{\mathrm{r}}^2} \right\}; \\
			
			\text{2)} R_{1,{\mathrm{N}}}^* = \max \{ {R_{{\mathrm{0}},{\mathrm{N}}}^*,R_{{\mathrm{1}},{\mathrm{N}}}^*,R_{{\mathrm{2}},{\mathrm{N}}}^*} \} \text{ if }
			d_{\mathrm{U}}^2 < \min \left\{ {{{\sigma _{\mathrm{r}}^2{P_{\mathrm{I}}}d_{\mathrm{B}}^2} \over {\sigma _0^2{P_{\mathrm{B}}}}},{{4\beta \beta _{\mathrm{I}}^2\beta _{\mathrm{B}}^2{P_{\mathrm{I}}}{P_{\mathrm{B}}}{N^3}} \over {27{c_1}{c_4}{{( {c_2^2/4 + N} )^2}}}}} \right\}; \\
			
			\text{3)} R_{2,{\mathrm{N}}}^* = \max \{ {R_{{\mathrm{0}},{\mathrm{N}}}^*,R_{{\mathrm{1}},{\mathrm{N}}}^*,R_{{\mathrm{2}},{\mathrm{N}}}^*} \} \text{ if } d_{\mathrm{B}}^2 < \min \left\{ {{{\sigma _0^2{P_{\mathrm{B}}}d_{\mathrm{U}}^2} \over {\sigma _{\mathrm{r}}^2{P_{\mathrm{I}}}}},{{4\beta \beta _{\mathrm{I}}^2\beta _{\mathrm{U}}^2{P_{\mathrm{I}}}{P_{\mathrm{B}}}{N^3}} \over {27{c_1}{c_8}{{( {c_2^2/4 + N} )^2}}}}} \right\}.
		\end{cases} \label{con_2}
	\end{equation}
	\vspace{-5pt}
	{\noindent} \rule[-0pt]{18.3cm}{0.05em}
\end{figure*}

\subsubsection{Comparison among Three Deployment Schemes}
For the elements allocation design, more passive elements should be deployed at the PIRS for both BAPU and BPAU scheme as the total number of IRS elements increases. However, it is different from the BHU scheme, which has been described in the analysis of Proposition \ref{pro_N_H} in detail. Considering the distance-dependent path loss model, let $\beta$ denote the the path loss at the reference distance, which is set to 1 meter (m). To facilitate performance comparison, the path-loss exponents of all the links are set to 2, i.e., $\beta _{{\rm{BI}}}^2 = {\beta }/{{d_{{\rm{BI}}}^2}}$, $\beta _{{\rm{IU}}}^2 = {\beta }/{{d_{{\rm{IU}}}^2}}$, $\beta _{{\rm{B}}}^2 = {\beta }/{{d_{{\rm{B}}}^2}}$, $\beta _{{\rm{I}}}^2 = {\beta }/{{d_{{\rm{I}}}^2}}$, and $\beta _{{\rm{U}}}^2 = {\beta }/{{d_{{\rm{U}}}^2}}$. Under the assumption of $\tilde{N}_{{\mathrm{p}}}^2 \ll \min \left\{ {c_4}/{c_3}, {c_8}/{c_7} \right\}$, the capacities for the BHU, BAPU, and BPAU schemes with low-complexity elements allocation solutions are given by $R_{0,{\mathrm{N}}}^* = {{\log _2}\left( {1 + {\gamma _\mathrm{0,N}^*}} \right)}$, $R_{1,{\mathrm{N}}}^* = {{\log _2}\left( {1 + {\gamma _\mathrm{1,N}^*}} \right)}$, and $R_{2,{\mathrm{N}}}^* = {{\log _2}\left( {1 + {\gamma _\mathrm{2,N}^*}} \right)}$, respectively. It is readily verified that the assumption is reasonable, especially when the total number of IRS elements is limited, since the two-hop path loss is generally significantly more severe than the single-hop path loss. In the following, we provide a theoretical comparison among the three IRS deployment schemes, which is expressed in \eqref{con_1} when $N \le {{{P_{\mathrm{I}}}}}/({{4{P_{\mathrm{B}}}\beta _{{\mathrm{BI}}}^2 + 4\sigma _{\mathrm{r}}^2}})$ and in \eqref{con_2} when $N > {{{P_{\mathrm{I}}}}/({4{P_{\mathrm{B}}}\beta _{{\mathrm{BI}}}^2 + 4\sigma _{\mathrm{r}}^2})}$. First, one can observe that the BHU performs better when deployed close to the user, i.e., $d_{{\mathrm{IU}}}$ is small. Second, we observe that the BAPU scheme is preferable when the PIRS is deployed close to the user, i.e., $d_{{\mathrm{U}}}$ is small. The reason is that stronger reflected signals are received at the user resulting in a higher SNR. Third, BPAU is preferable if the PIRS is deployed closer to the BS, i.e., a smaller $d_{{\mathrm{B}}}$ since the power of the incident signal at the PIRS from the BS is stronger. The results highlight the importance of carefully designing the IRS deployment, which will be further investigated in the next sub-section.

\subsection{IRS Placement Optimization}
For any given elements allocation, we next optimize the IRS placement. Based on \eqref{SNR_H_cf}, we formulate the optimization problem for BHU scheme as follows
\begin{subequations}
	\label{Problem_placement_1}
	\begin{align}
		\mathop {\max }\limits_{x_\mathrm{BI},x_\mathrm{IU}} \;\; &\gamma _0\\ \;\;\;\;\;\;\;\; {\mathrm{s.t.}} \;\;\;\; & x_\mathrm{BI} + x_\mathrm{IU} = L, \\
		& x_\mathrm{BI} \ge 0, x_\mathrm{IU} \ge 0. 
	\end{align}
\end{subequations}
Based on \eqref{SNR_AP_cf} and \eqref{SNR_PA_cf}, for the case with BAPU scheme ($i = \mathrm{1}$) or BPAU scheme ($i = \mathrm{2}$), the optimization problem can be rewritten as
\begin{subequations}
	\label{Problem_placement_2}
	\begin{align}
		\mathop {\max }\limits_{x_\mathrm{B},x_\mathrm{I},x_\mathrm{U}} \;\;\; &\gamma _i \\{\mathrm{s.t.}} \;\;\;\;\; &i \in \left\{1,2\right\}, \\
		& x_\mathrm{B} + x_\mathrm{I} + x_\mathrm{U} = L, \\
		& x_\mathrm{B} \ge 0, x_\mathrm{I} \ge 0, x_\mathrm{U} \ge 0.
	\end{align}
\end{subequations}
Although the optimal solution to problem \eqref{Problem_placement_1} and problem \eqref{Problem_placement_2} can be obtained through a one-dimensional search over $x_\mathrm{BI} \in \left[0,L\right]$ with $x_\mathrm{IU} = L - x_\mathrm{BI}$ and a two-dimensional search over $x_\mathrm{B} \in \left[0,L\right]$ and $x_\mathrm{I} \in \left[0,L - x_\mathrm{B} \right]$, respectively, it yields little useful insights on the IRS placement. In the following, we first derive the high-quality sub-optimal solution to the above problems and then compare the performance of the systems with three IRS deployment schemes. To facilitate derivation, we set the path-loss exponents for the considered links as $\alpha_{\mathrm{BI}} = \alpha_{\mathrm{IU}} = \alpha_{\mathrm{B}} = \alpha_{\mathrm{I}} = \alpha_{\mathrm{U}} = 2$. Moreover, we have $\beta \sigma _{\mathrm{r}}^2{P_{\mathrm{I}}}d_{{\mathrm{BI}}}^2 + \beta \sigma _0^2{P_{\mathrm{B}}}d_{{\mathrm{IU}}}^2 \gg \sigma _{\mathrm{r}}^2\sigma _0^2d_{{\mathrm{IU}}}^2d_{{\mathrm{BI}}}^2$, ${\beta ^2}\sigma _{\mathrm{r}}^2{P_{\mathrm{I}}}N_{{\mathrm{p}}}^2d_{\mathrm{B}}^2 + \beta \sigma _0^2{P_{\mathrm{B}}}d_{\mathrm{I}}^2d_{\mathrm{U}}^2 \gg \sigma _0^2\sigma _{\mathrm{r}}^2d_{\mathrm{B}}^2d_{\mathrm{I}}^2d_{\mathrm{U}}^2$, and $\beta \sigma _{\mathrm{r}}^2{P_{\mathrm{I}}}d_{\mathrm{I}}^2d_{\mathrm{B}}^2 + {\beta ^2}\sigma _0^2{P_{\mathrm{B}}}N_{{\mathrm{p}}}^2d_{\mathrm{U}}^2 \gg \sigma _0^2\sigma _{\mathrm{r}}^2d_{\mathrm{B}}^2d_{\mathrm{I}}^2d_{\mathrm{U}}^2$ if \eqref{con_x} holds. The above assumptions are reasonable for the following reasons. First, the two-hop path loss is generally significantly more severe than the single-hop path loss. Second, the value of the noise power is typically small without loss of generality.
\begin{figure*}[ht]
	\begin{align}
		\label{con_x}
		{P_{\mathrm{B}}} \gg \max \left\{ \frac{{\sigma _{\mathrm{r}}^2d_{{\mathrm{BI}}}^2}}{\beta } - \frac{{\sigma _{\mathrm{r}}^2{P_{\mathrm{I}}}d_{{\mathrm{BI}}}^2}}{{\sigma _0^2d_{{\mathrm{IU}}}^2}}, \frac{{d_{\mathrm{B}}^2\sigma _{\mathrm{r}}^2}}{\beta } - \frac{{\sigma _{\mathrm{r}}^2\beta d_{\mathrm{B}}^2{P_{\mathrm{I}}}N_{{\mathrm{p}}}^2}}{{\sigma _0^2d_{\mathrm{I}}^2d_{\mathrm{U}}^2}}, \frac{{\sigma _{\mathrm{r}}^2d_{\mathrm{I}}^2d_{\mathrm{B}}^2}}{{{\beta ^2}N_{{\mathrm{p}}}^2}} - \frac{{\sigma _{\mathrm{r}}^2{P_{\mathrm{I}}}d_{\mathrm{I}}^2d_{\mathrm{B}}^2}}{{\beta \sigma _0^2N_{{\mathrm{p}}}^2d_{\mathrm{U}}^2}} \right\}.
	\end{align}
	{\noindent} \rule[-0pt]{18.3cm}{0.05em}
	\vspace{-20pt}
\end{figure*}

\subsubsection{BHU Scheme}
\label{sec_x_h}
Under the assumption of $\beta \sigma _{\mathrm{r}}^2{P_{\mathrm{I}}}d_{{\mathrm{BI}}}^2 + \beta \sigma _0^2{P_{\mathrm{B}}}d_{{\mathrm{IU}}}^2 \gg \sigma _{\mathrm{r}}^2\sigma _0^2d_{{\mathrm{IU}}}^2d_{{\mathrm{BI}}}^2$, the SNR is approximated as ${\gamma _{{\mathrm{0,x}}}} (x_\mathrm{BI}) = \frac{{{P_{\mathrm{B}}}\beta {{\left( {\sqrt {{N_{{\mathrm{a}}}}{P_{\mathrm{I}}}}  + \sqrt {{N_{{\mathrm{p}}}^2 {P_{\mathrm{B}}}\beta /( {x_\mathrm{BI}^2 + h_\mathrm{s}^2} ) }} } \right)^2}}}}{{\sigma _{\mathrm{r}}^2{P_{\mathrm{I}}} ( {x_\mathrm{BI}^2 + h_\mathrm{s}^2} ) + \sigma _0^2{P_{\mathrm{B}}}{\left({( {L - {x_\mathrm{BI}}} )^2}+h_\mathrm{s}^2\right)}}}$. Thus, problem \eqref{Problem_placement_1} can be reformulated as
\begin{subequations}
	\label{Problem_placement_1_1}
	\begin{align}
		\mathop {\max }\limits_{x_\mathrm{BI}} \;\;\; &{\gamma _{{\mathrm{0,x}}} \left(x_\mathrm{BI}\right)} \\ {\mathrm{s.t.}} \;\;\; & 0 \le x_\mathrm{BI} \le L.
	\end{align}
\end{subequations}

\begin{Proposition}
	\label{pro_x}
	In the high SNR case, the solution to problem \eqref{Problem_placement_1_1} is given by
	\begin{align}
		\label{x_h}
		x_\mathrm{BI}^* = \frac{{{\sigma _0^2{P_{\mathrm{B}}}}L}}{{{\sigma _0^2{P_{\mathrm{B}}}} + {\sigma _{\mathrm{r}}^2{P_{\mathrm{I}}}}}}.
	\end{align}
	The SNR is approximated as
	\begin{align}
		\label{SNR_0_x}
		\gamma _{{\mathrm{0}},{\mathrm{x}}}^* \!\!=\!\! \frac{{( {{c_9} \!+\! {c_{10}}} ){P_{\mathrm{B}}}\beta {{\left(\!\!\! {\sqrt {{N_{{\mathrm{a}}}}{P_{\mathrm{I}}}}  \!+\! \sqrt {\frac{{N_{{\mathrm{p}}}^2{P_{\mathrm{B}}}\beta {{( {{c_9} + {c_{10}}} )^2}}}}{{c_9^2{L^2} + h_\mathrm{s}^2{{( {{c_9} + {c_{10}}} )^2}}}}} } \right)^2}}}}{{h_\mathrm{s}^2{{( {{c_9} + {c_{10}}} )^2}} + {c_9}{c_{10}}{L^2}}},
	\end{align}
	with $c_9 = {\sigma _0^2{P_{\mathrm{B}}}}$ and $c_{10} = {\sigma _{\mathrm{r}}^2{P_{\mathrm{I}}}}$.
\end{Proposition}

{\it{Proof:}}
Please refer to Appendix C.
~$\hfill\blacksquare$

Proposition \eqref{pro_x} shows that the optimized BS-HIRS distance is independent of the number of passive or active reflecting elements, i.e., $N_\mathrm{p}$ and $N_\mathrm{a}$, but is determined by the transmit power and amplification power. It is practically appealing because both IRS elements allocation and placement are obtained in closed-form expressions and there is no need for multiple alternating iterations between them. Moreover, they are jointly optimized rather than completely independently. Specifically, with the optimized IRS position, we then optimize the IRS elements allocation. This approach reduces computational complexity while ensuring effective results. As such, the SNR with the design of optimized IRS elements allocation and placement for the BHU scheme can be expressed as \eqref{y_0_opt} on the top of next page.
\begin{figure*}
\begin{align}
	\label{y_0_opt}
	\gamma _{{\mathrm{0}}}^* = \left\{ {\begin{array}{*{20}{c}}
			{\frac{{( {{c_9} + {c_{10}}} ){P_{\mathrm{B}}}\beta {{\left( {\sqrt {{\left(N-1\right)}{P_{\mathrm{I}}}}  + \sqrt {\frac{{{P_{\mathrm{B}}}\beta {{( {{c_9} + {c_{10}}} )^2}}}}{{c_9^2{L^2} + h_\mathrm{s}^2{{( {{c_9} + {c_{10}}} )^2}}}}} } \right)^2}}}}{{h_\mathrm{s}^2{{( {{c_9} + {c_{10}}} )^2}} + {c_9}{c_{10}}{L^2}}},}&{{{N \le \frac{{{P_{\mathrm{I}}}}}{{4 {{P_{\mathrm{B}}}\beta _{{\mathrm{BI}}}^2 + 4 \sigma _{\mathrm{r}}^2} }}}},}\\
			{\mathop {\max }\limits_{{N_{\rm{a}}}} {\gamma _{0,{N_{\rm{a}}}}}\left( {{N_{\rm{a}}}} \right),{N_{\rm{a}}} \in \left\{ {\left\lfloor {\frac{{{P_{\rm{I}}}}}{{4{P_{\rm{B}}}\beta _{{\rm{BI}}}^2 + 4\sigma _{\rm{r}}^2}}} \right\rfloor ,\left\lceil {\frac{{{P_{\rm{I}}}}}{{4{P_{\rm{B}}}\beta _{{\rm{BI}}}^2 + 4\sigma _{\rm{r}}^2}}} \right\rceil } \right\},}&{\text{Otherwise.}}
	\end{array}} \right.
\end{align}
\vspace{-10pt}
{\noindent} \rule[-0pt]{18.3cm}{0.05em}
\end{figure*}

\subsubsection{BAPU Scheme}
\label{sec_x_ap}
Note that placing a single PIRS close to the user/AP yields a larger SNR \cite{survey_x1}. For the BAPU scheme, we consider a scenario where the PIRS is deployed directly above the user, i.e., $x_{\mathrm{U}} = 0$, thereby focusing on the deployment of AIRS. Under the assumption of ${\beta ^2}\sigma _{\mathrm{r}}^2{P_{\mathrm{I}}}N_{{\mathrm{p}}}^2d_{\mathrm{B}}^2 + \beta \sigma _0^2{P_{\mathrm{B}}}d_{\mathrm{I}}^2d_{\mathrm{U}}^2 \gg \sigma _0^2\sigma _{\mathrm{r}}^2d_{\mathrm{B}}^2d_{\mathrm{I}}^2d_{\mathrm{U}}^2$, problem \eqref{Problem_placement_2} with $i=1$ can be reformulated as
\begin{subequations}
	\label{Problem_placement_2_1}
	\begin{align}
		\mathop {\min }\limits_{x_{\mathrm{B}}} \;\;\; &{\beta}\sigma _{\mathrm{r}}^2{P_{\mathrm{I}}}N_{{\mathrm{p}}}^2 \left({x_{\mathrm{B}}^2 + h_\mathrm{d}^2}\right) + \sigma _0^2{P_{\mathrm{B}}} h_\mathrm{d}^2 {{\left(L - x_{\mathrm{B}}\right)^2}}\\ {\mathrm{s.t.}} \;\;\; &0 \le {x_{\mathrm{B}}} \le L.
	\end{align}
\end{subequations}
By setting the first-order partial derivative of the objective function to zero, the solution to problem \eqref{Problem_placement_2_1} is given by
\begin{align}
	\label{x_ap}
	x_{\mathrm{B}}^* = \frac{L \sigma _0^2{P_{\mathrm{B}}} h_\mathrm{d}^2}{{\beta}\sigma _{\mathrm{r}}^2{P_{\mathrm{I}}}N_{{\mathrm{p}}}^2 + \sigma _0^2{P_{\mathrm{B}}} h_\mathrm{d}^2}.
\end{align}
As such, the SNR is approximated as
\begin{align}
	\label{SNR_1_x}
	\gamma _{{\mathrm{1,x}}}^* = \frac{{\beta {P_{\mathrm{B}}}{N_{{\mathrm{a}}}}\left( {\beta {c_{10}}N_{{\mathrm{p}}}^2 + {c_9}h_{\mathrm{d}}^2} \right)}}{{\sigma _{\mathrm{r}}^2h_{\mathrm{d}}^2\left( {\beta {c_{10}}N_{{\mathrm{p}}}^2 + {c_9}{L^2}} \right)}}.
\end{align}

\subsubsection{BPAU Scheme}
\label{sec_x_pa}
Under the assumption of $\beta \sigma _{\mathrm{r}}^2{P_{\mathrm{I}}}d_{\mathrm{I}}^2d_{\mathrm{B}}^2 + {\beta ^2}\sigma _0^2{P_{\mathrm{B}}}N_{{\mathrm{p}}}^2d_{\mathrm{U}}^2 \gg \sigma _0^2\sigma _{\mathrm{r}}^2d_{\mathrm{B}}^2d_{\mathrm{I}}^2d_{\mathrm{U}}^2$, we focus on the case where the PIRS is placed directly above the BS, i.e., $x_{\mathrm{B}} = 0$ and reformulate problem \eqref{Problem_placement_2} with $i=2$ as
\begin{subequations}
	\label{Problem_placement_2_2}
	\begin{align}
		\mathop {\min }\limits_{x_{\mathrm{U}}} \;\;\; &{{\beta}\sigma _0^2{P_{\mathrm{B}}}N_{{\mathrm{p}}}^2} \left(x_{\mathrm{U}}^2+h_\mathrm{d}^2\right) + { \sigma _{\mathrm{r}}^2{P_{\mathrm{I}}}} h_\mathrm{d}^2{{\left( {L - x_{\mathrm{U}}} \right)^2}}\\ {\mathrm{s.t.}} \;\;\; &0 \le x_{\mathrm{U}} \le L.
	\end{align}
\end{subequations}
By setting the first-order partial derivative of the objective function to zero, the solution to problem \eqref{Problem_placement_2_2} is given by 
\begin{align}
	\label{x_pa}
	x_{\mathrm{U}}^* = \frac{L {\sigma _{\mathrm{r}}^2{P_{\mathrm{I}}}} h_\mathrm{d}^2}{{\beta \sigma _0^2{P_{\mathrm{B}}}N_{{\mathrm{p}}}^2} + {\sigma _{\mathrm{r}}^2{P_{\mathrm{I}}}} h_\mathrm{d}^2}.
\end{align}
As such, the SNR is approximated as
\begin{align}
	\label{SNR_2_x}
	\gamma _{{\mathrm{2,x}}}^* = \frac{{\beta {P_{\mathrm{I}}}{N_{{\mathrm{a}}}}\left( {\beta {c_9}N_{{\mathrm{p}}}^2 + {c_{10}}h_{\mathrm{d}}^2} \right)}}{{\sigma _0^2h_{\mathrm{d}}^2\left( {\beta {c_9}N_{{\mathrm{p}}}^2 + {c_{10}}{L^2}} \right)}}.
\end{align}

\subsubsection{Comparison among Three Deployment Schemes} In the following, we first characterize the impact of system parameters on the optimized BS-HIRS/AIRS horizontal distance and then we provide a theoretical comparison in terms of SNR among the three schemes with IRS placement optimization. 

One can observe from \eqref{x_h}, \eqref{x_ap}, and \eqref{x_pa} that the optimized horizontal distance between the BS and the HIRS/AIRS monotonically increases with $P_{\mathrm{B}}$ and decreases with $P_{\mathrm{I}}$ under the three schemes. With a larger transmit power $P_{\mathrm{B}}$, the HIRS/AIRS moves towards the user to provide a higher amplification gain and reduce the path loss in the next hop, thus enhancing the signal strength received by the user. As $P_{\mathrm{I}}$ increases, the HIRS/AIRS should be deployed closer to BS to minimize noise amplification, thereby suppressing the noise power received at the user. Moreover, we observe that the optimized IRS deployment for both BAPU and BPAU schemes depends on the number of passive elements $N_\mathrm{p}$, whereas that for BHU schemes is independent of $N_\mathrm{p}$. 

\begin{figure*}[ht]
	\small
	\begin{align}
		\frac{{\gamma _{{\rm{1}},{\rm{x}}}^*}}{{\gamma _{{\rm{0}},{\rm{x}}}^*}} =& \frac{{{N_{{\rm{act}}}}\left( {\beta {c_{10}}N_{{\rm{pas}}}^2 + {c_9}h_{\rm{d}}^2} \right)\left( {h_{\rm{s}}^2{{\left( {{c_9} + {c_{10}}} \right)}^2} + {c_9}{c_{10}}{L^2}} \right)}}{{\sigma _{\rm{r}}^2h_{\rm{d}}^2\left( {\beta {c_{10}}N_{{\rm{pas}}}^2 + {c_9}{L^2}} \right)\left( {{c_9} + {c_{10}}} \right){{\left( {\sqrt {{N_{{\rm{act}}}}{P_{\rm{I}}}}  + \sqrt {N_{{\rm{pas}}}^2{P_{\rm{B}}}\beta {{\left( {{c_9} + {c_{10}}} \right)}^2}/\left( {c_9^2{L^2} + h_{\rm{s}}^2{{\left( {{c_9} + {c_{10}}} \right)}^2}} \right)} } \right)}^2}}}, \label{SNR_AP_H} \\
		\frac{{\gamma _{{\rm{2}},{\rm{x}}}^*}}{{\gamma _{{\rm{0}},{\rm{x}}}^*}} =& \frac{{{P_{\rm{I}}}{N_{{\rm{act}}}}\left( {\beta {c_9}N_{{\rm{pas}}}^2 + {c_{10}}h_{\rm{d}}^2} \right)\left( {h_{\rm{s}}^2{{\left( {{c_9} + {c_{10}}} \right)}^2} + {c_9}{c_{10}}{L^2}} \right)}}{{{P_{\rm{B}}}\sigma _0^2h_{\rm{d}}^2\left( {\beta {c_9}N_{{\rm{pas}}}^2 + {c_{10}}{L^2}} \right)\left( {{c_9} + {c_{10}}} \right){{\left( {\sqrt {{N_{{\rm{act}}}}{P_{\rm{I}}}}  + \sqrt {N_{{\rm{pas}}}^2{P_{\rm{B}}}\beta {{\left( {{c_9} + {c_{10}}} \right)}^2}/\left( {c_9^2{L^2} + h_{\rm{s}}^2{{\left( {{c_9} + {c_{10}}} \right)}^2}} \right)} } \right)}^2}}}. \label{SNR_PA_H}
	\end{align}
	\vspace{-10pt}
	{\noindent} \rule[-0pt]{18.3cm}{0.05em}
\end{figure*}
\begin{figure*}
	\begin{align}
		\label{con_x_ap_pa}
		( {{c_9} - {c_{10}}} ) ( ( {( {\beta N_{{\mathrm{p}}}^2 - {L^2}} )\beta N_{{\mathrm{p}}}^2 + ( {\beta N_{{\mathrm{p}}}^2 + {L^2}} )h_{\mathrm{d}}^2} ){c_9}{c_{10}} + ( {c_9^2 + c_{10}^2} )\beta N_{{\mathrm{p}}}^2h_{\mathrm{d}}^2) > 0.
	\end{align}
	\vspace{-10pt}
	{\noindent} \rule[-0pt]{18.3cm}{0.05em}
\end{figure*}
Based on \eqref{SNR_0_x}, \eqref{SNR_1_x} and \eqref{SNR_2_x}, the SNR ratio of BAPU scheme to BHU scheme and BPAU scheme to BHU scheme are expressed as \eqref{SNR_AP_H} and \eqref{SNR_PA_H} on the top of this page, respectively. As ${P_{\mathrm{B}}} \to 0$, the SNR ratio is bounded by ${{\gamma _{{\rm{1}},{\rm{x}}}^*}}/{{\gamma _{{\rm{0}},{\rm{x}}}^*}} \to {{h_{\rm{s}}^2}}/{{h_{\rm{d}}^2}}$. Generally, HIRS should be placed higher to cover the BS and user, i.e., ${{h_{\rm{d}}}}>{{h_{\rm{s}}^2}}$. It reveals that the BHU scheme with IRS placement optimization outperforms the BAPU scheme when the transmit power is insufficient. As ${P_{\mathrm{I}}} \to 0$, it follows that ${{\gamma _{{\rm{2}},{\rm{x}}}^*}}/{{\gamma _{{\rm{0}},{\rm{x}}}^*}} \to 0$, which shows that BHU performs better than BPAU with a small amplification power. The reason is that the rate performance is mainly limited by IRS 2, i.e., the PIRS in BAPU scheme and the AIRS in BPAU scheme, which results in a weak received signal at the user.

Next, we further compare the BAPU and BPAU schemes theoretically. Given $L \ge h_{\mathrm{d}}$, it can be readily verified from \eqref{SNR_1_x} and \eqref{SNR_2_x} that the SNR for both BAPU and BPAU scheme increases with $P_{\mathrm{B}}$ and $P_{\mathrm{I}}$, respectively. However, the increase of SNR for the BAPU scheme with optimized placement is more sensitive to the transmit power, rather than the amplification power as that with optimized elements allocation. In contrast, the increase of SNR for the BPAU scheme is more sensitive to the amplification power. Moreover, BAPU outperforms BPAU when \eqref{con_x_ap_pa} holds. In the general case with ${c_9} > {c_{10}}$ and $\beta N_{{\mathrm{p}}}^2 - {L^2} < 0$, BPAU performs better than BAPU if $h_{\mathrm{d}}^2 < \frac{{ \left( {{L^2} - \beta N_{{\mathrm{p}}}^2} \right)\beta N_{{\mathrm{p}}}^2{c_9}{c_{10}}}}{{\left( {\beta N_{{\mathrm{p}}}^2 + {L^2}} \right){c_9}{c_{10}} + \left( {c_9^2 + c_{10}^2} \right)\beta N_{{\mathrm{p}}}^2}}$ because it suffers less path loss in the first hop and thus results in higher received power despite the small number of passive elements. Furthermore, both schemes achieve the same SNR if $\sigma _0^2{P_{\mathrm{B}}} = \sigma _{\mathrm{r}}^2{P_{\mathrm{I}}}$.

\vspace{-5pt}
\section{Asymptotic Performance Analysis}
\label{order}
Note that the results in Section \ref{EAP_opt} are applicable to any given $N$, ${P_{\mathrm{B}}}$, and ${P_{\mathrm{I}}}$. To obtain more insights, we provide asymptotic analysis on the BHU, BAPU, and BPAU schemes when $N$, ${P_{\mathrm{B}}}$, or ${P_{\mathrm{I}}}$ is sufficiently large.
\subsection{Asymptotic SNR and Capacity Characterization}
In this subsection, we first obtain the asymptotic SNR for three deployment schemes w.r.t. $N$, ${P_{\mathrm{B}}}$, and ${P_{\mathrm{I}}}$. Based on the above, we characterize the corresponding system capacity scaling orders. It is worth noting that the capacity scaling orders are independent of IRS elements allocation and placement because they have no impact on the asymptotic result.
\subsubsection{BHU Scheme}
Let $N_\mathrm{p} = \varepsilon N$. Then, we have $N_\mathrm{a} = (1 - \varepsilon) N$. Based on \eqref{SNR_H_cf}, as ${N \to \infty }$, the asymptotic SNR is given by
\begin{align}
	\label{sym_SNR_0_N}
	{\gamma _0} \to {N^2}\frac{{{P_{\mathrm{B}}}\beta _{{\mathrm{BI}}}^2\beta _{{\mathrm{IU}}}^2{\varepsilon ^2}\left( {{P_{\mathrm{B}}}\beta _{{\mathrm{BI}}}^2 + \sigma _{\mathrm{r}}^2} \right)}}{{\sigma _{\mathrm{r}}^2{P_{\mathrm{I}}}\beta _{{\mathrm{IU}}}^2 + \sigma _0^2{P_{\mathrm{B}}}\beta _{{\mathrm{BI}}}^2 + \sigma _0^2\sigma _{\mathrm{r}}^2}}.
\end{align}
As ${{P_{\mathrm{B}}} \to \infty }$, it follows that 
\begin{align}
	\label{sym_SNR_0_Pb}
	{\gamma _0} \to {P_{\mathrm{B}}}\frac{{2\beta _{{\mathrm{BI}}}^2\beta _{{\mathrm{IU}}}^2N_{{\mathrm{p}}}^2}}{{\sigma _0^2}}.
\end{align}
As ${{P_{\mathrm{I}}} \to \infty }$, the asymptotic SNR is upper-bounded by
\begin{align}
	\label{sym_SNR_0_Pi}
	{\gamma _0} \to {{{P_{\mathrm{B}}}\beta _{{\mathrm{BI}}}^2{N_{{\mathrm{a}}}}}}/{{\sigma _{\mathrm{r}}^2}}.
\end{align}
One can observe from \eqref{sym_SNR_0_N} and \eqref{sym_SNR_0_Pi} that the asymptotic SNR depends on the BS-HIRS channel gain and the noise power at the active elements as ${N \to \infty }$ or ${{P_{\mathrm{I}}} \to \infty }$, which indicates that the HIRS should be placed closer to the BS with a large total number of IRS elements $N$ or sufficient amplification power of the active elements ${P_{\mathrm{I}}}$. From \eqref{sym_SNR_0_Pb}, it is observed that the asymptotic SNR is independent of $\sigma _{\mathrm{r}}^2$, while it depends on the number of passive elements and the channel gains of all involved links, which reveals that increasing the BS transmit power ${P_{\mathrm{B}}}$ can reduce the negative impact caused by a large $\sigma _{\mathrm{r}}^2$. As ${{P_{\mathrm{I}}} \to \infty }$, the asymptotic SNR increases with $N_\mathrm{a}$, which shows that more active elements should be deployed at the HIRS given sufficient amplification power.

\subsubsection{BAPU Scheme}
Based on \eqref{SNR_AP_cf}, as ${N \to \infty }$, it follows that
\begin{align}
	\label{sym_SNR_1_N}
	{\gamma _1} \to N\frac{{3{P_{\mathrm{B}}}\beta _{\mathrm{B}}^2\left( {1 - \varepsilon } \right)}}{{\sigma _{\mathrm{r}}^2}}.
\end{align}
In contrast to the BHU scheme where the asymptotic SNR is proportional to ${P_{\mathrm{B}}}$ as ${{P_{\mathrm{B}}} \to \infty }$, the BAPU scheme exhibits an upper bound for the asymptotic SNR denoted as
\begin{align}
	\label{sym_SNR_1_Pb}
	{\gamma _1} \to {{{\beta _{{\mathrm{U}}}^2\beta _{{\mathrm{I}}}^2}{N_{{\mathrm{a}}}}N_{{\mathrm{p}}}^2{P_{\mathrm{I}}}}}/{{\sigma _0^2}}.
\end{align}
As ${{P_{\mathrm{I}}} \to \infty }$, the asymptotic SNR will be upper-bounded by
\begin{align}
	\label{sym_SNR_1_Pi}
	{\gamma _1} \to {{{\beta _{{\mathrm{B}}}^2}{N_{{\mathrm{a}}}}{P_{\mathrm{B}}}}}/{{\sigma _{\mathrm{r}}^2}}.
\end{align}
Similar to the BHU scheme, the AIRS should be placed closer to the BS for BAPU scheme when the total number of IRS elements or amplification power budget of the AIRS is sufficiently large. From \eqref{sym_SNR_1_Pb}, we observe that the asymptotic SNR is depends on the channel gains of the AIRS→PIRS→user link rather than the entire link from BS to user as in the BHU scheme. The impact of $\sigma _{\mathrm{r}}^2$ and $N_{{\mathrm{a}}}$ on the asymptotic SNR as ${P_{\mathrm{B}}} \to \infty$ and ${P_{\mathrm{I}}} \to \infty$, respectively, are the same as in the BHU scheme, for which the explanation is omitted for brevity. 

\subsubsection{BPAU Scheme}
Based on \eqref{SNR_PA_cf}, as ${N \to \infty }$, we have
\begin{align}
	\label{sym_SNR_2_N}
	{\gamma _2} \to N\frac{{3\beta _{\mathrm{U}}^2\left( {1 - \varepsilon } \right){P_{\mathrm{I}}}}}{{\sigma _0^2}}.
\end{align}
As ${{P_{\mathrm{B}}} \to \infty }$, the upper bound of the asymptotic SNR is given by 
\begin{align}
	\label{sym_SNR_2_Pb}
	{\gamma _2} \to {{\beta _{\mathrm{U}}^2{N_{{\mathrm{a}}}}{P_{\mathrm{I}}}}}/{{\sigma _0^2}}.
\end{align}
As ${{P_{\mathrm{I}}} \to \infty }$, the asymptotic SNR is upper-bounded by
\begin{align}
	\label{sym_SNR_2_Pi}
	{\gamma _2} \to {{{P_{\mathrm{B}}}\beta _{\mathrm{I}}^2\beta _{\mathrm{B}}^2{N_{{\mathrm{a}}}}N_{{\mathrm{p}}}^2}}/{{\sigma _{\mathrm{r}}^2}}.
\end{align}  
From \eqref{sym_SNR_2_N} and \eqref{sym_SNR_2_Pb}, we observe that the asymptotic SNR depends on the AIRS-user channel gain and is independent of the noise power at the AIRS as ${N \to \infty }$ or ${{P_{\mathrm{B}}} \to \infty }$. It unveils that the AIRS should be placed closer to the user when the total number of IRS elements or the BS transmit power is sufficiently large. Different from the previous two schemes, the results unveil that the negative impact of large $\sigma _{\mathrm{r}}^2$ can be alleviated not only by increasing the BS transmit power but also by deploying more IRS elements under the BPAU scheme. Moreover, it is observed from \eqref{sym_SNR_2_Pi} that the upper bound of the asymptotic SNR increases with both the number of active elements and passive elements as ${{P_{\mathrm{I}}} \to \infty }$.

\subsection{Performance Comparison}
\subsubsection{Finite-Scale Analysis}
To determine the best of the three considered schemes, we begin by comparing the SNR of BAPU and BPAU schemes, followed by a comparison with BHU scheme. Based on \eqref{SNR_AP_cf} and \eqref{SNR_PA_cf}, the SNR ratio of BAPU scheme to BPAU scheme is given by
\begin{align}
	\label{SNR_double}
	\frac{{{\gamma _1}}}{{{\gamma _2}}} = \frac{{\sigma _0^2{P_{\mathrm{B}}}\beta _{\mathrm{I}}^2\beta _{\mathrm{B}}^2N_{{\mathrm{p}}}^2 + \beta _{\mathrm{U}}^2\sigma _{\mathrm{r}}^2{P_{\mathrm{I}}} + \sigma _0^2\sigma _{\mathrm{r}}^2}}{{\beta _{\mathrm{U}}^2\beta _{\mathrm{I}}^2\sigma _{\mathrm{r}}^2{P_{\mathrm{I}}}N_{{\mathrm{p}}}^2 + \sigma _0^2{P_{\mathrm{B}}}\beta _{\mathrm{B}}^2 + \sigma _0^2\sigma _{\mathrm{r}}^2}}.
\end{align}
From \eqref{SNR_double}, it is observed that BAPU outperforms BPAU if and only if
\begin{align}
	\label{condition_double}
	\left( {{P_{\mathrm{I}}}\beta _{\mathrm{U}}^2\sigma _{\mathrm{r}}^2 - {P_{\mathrm{B}}}\beta _{\mathrm{B}}^2\sigma _0^2} \right)\left( {1 - {N_{{\mathrm{p}}}^2}\beta _{\mathrm{I}}^2} \right) > 0.
\end{align}

\begin{figure*}[t]
	\vspace{-10pt}
	\begin{align}
		\label{SNR_H_AP}
		\frac{{{\gamma _0}}}{{{\gamma _1}}} = \frac{{\beta _{{\rm{BI}}}^2\beta _{{\rm{IU}}}^2{{\left( {\sqrt {{N_{{\rm{act}}}}{P_{\rm{I}}}}  + \sqrt {N_{{\rm{pas}}}^2\left( {{P_{\rm{B}}}\beta _{{\rm{BI}}}^2 + \sigma _{\rm{r}}^2} \right)} } \right)}^2}\left( {N_{{\rm{pas}}}^2\beta _{\rm{U}}^2\beta _{\rm{I}}^2\sigma _{\rm{r}}^2{P_{\rm{I}}} + \sigma _0^2{P_{\rm{B}}}\beta _{\rm{B}}^2 + \sigma _0^2\sigma _{\rm{r}}^2} \right)}}{{\beta _{\rm{B}}^2\beta _{\rm{I}}^2\beta _{\rm{U}}^2{N_{{\rm{act}}}}N_{{\rm{pas}}}^2{P_{\rm{I}}}\left( {\sigma _{\rm{r}}^2{P_{\rm{I}}}\beta _{{\rm{IU}}}^2 + \sigma _0^2{P_{\rm{B}}}\beta _{{\rm{BI}}}^2 + \sigma _0^2\sigma _{\rm{r}}^2} \right)}},
	\end{align}
	\begin{align}
		\label{SNR_H_PA}
		\frac{{{\gamma _0}}}{{{\gamma _2}}} = \frac{{\beta _{{\rm{BI}}}^2\beta _{{\rm{IU}}}^2{{\left( {\sqrt {{N_{{\rm{act}}}}{P_{\rm{I}}}}  + \sqrt {N_{{\rm{pas}}}^2\left( {{P_{\rm{B}}}\beta _{{\rm{BI}}}^2 + \sigma _{\rm{r}}^2} \right)} } \right)}^2}\left( {\beta _{\rm{U}}^2\sigma _{\rm{r}}^2{P_{\rm{I}}} + \sigma _0^2{P_{\rm{B}}}\beta _{\rm{I}}^2\beta _{\rm{B}}^2N_{{\rm{pas}}}^2 + \sigma _0^2\sigma _{\rm{r}}^2} \right)}}{{\beta _{\rm{B}}^2\beta _{\rm{I}}^2\beta _{\rm{U}}^2{N_{{\rm{act}}}}N_{{\rm{pas}}}^2{P_{\rm{I}}}\left( {\sigma _{\rm{r}}^2{P_{\rm{I}}}\beta _{{\rm{IU}}}^2 + \sigma _0^2{P_{\rm{B}}}\beta _{{\rm{BI}}}^2 + \sigma _0^2\sigma _{\rm{r}}^2} \right)}}.
	\end{align}
	\vspace{-15pt}
	{\noindent} \rule[-0pt]{18.3cm}{0.05em}
\end{figure*}
In the large-${P_{\mathrm{I}}}$ regime, \eqref{condition_double} holds if ${N_{{\mathrm{p}}}^2}\beta _{\mathrm{I}}^2 > 1$. It means that the attenuation from path loss $\beta _{\mathrm{I}}^2$ can be compensated by the beamforming gain provided by PIRS in the BAPU scheme. In the large-${P_{\mathrm{B}}}$ regime, \eqref{condition_double} is easier to hold for small number of passive reflecting elements. This is expected since deploying active elements on the IRS 1 can amplify the attenuated signal after the first hop, thereby overcoming the path loss attenuation and effectively improving the SNR of the system. Based on \eqref{SNR_H_cf}, \eqref{SNR_AP_cf}, and \eqref{SNR_PA_cf}, the SNR ratio of BHU scheme to BAPU scheme and BHU scheme to BPAU scheme are expressed as \eqref{SNR_H_AP} and \eqref{SNR_H_PA} on the top of next page, respectively. However, it is challenging to obtain intuitive insights due to the complication expression. Thus, we further compare the three schemes through asymptotic results.

\subsubsection{Asymptotic Analysis}
\begin{table}[t]
	\centering
	\caption{Comparison of Capacity Scaling Order w.r.t. $N$, ${P_{\mathrm{B}}}$ and ${P_{\mathrm{I}}}$}
	\label{comp_R}
	\renewcommand\arraystretch{1.3}
	\begin{tabular}{|l|l|l|l|}
		\hline
		Scheme & ${N \to \infty }$ & ${{P_{\mathrm{B}}} \to \infty } $ & ${{P_{\mathrm{I}}} \to \infty }$ \\ \hline
		BHU ($R_0$) & ${\mathcal{O}(N^2)}$ & ${\mathcal{O}({P_{\mathrm{B}}})}$ & constant   \\ \hline
		BAPU ($R_1$) & ${\mathcal{O}(N)}$ & constant   & constant   \\ \hline
		BPAU ($R_2$) & ${\mathcal{O}(N)}$ & constant   & constant   \\ \hline
	\end{tabular}
	\vspace{-10pt}
\end{table}
Based on the asymptotic SNR for the three schemes in the previous subsection, it is not difficult to derive the corresponding characterize capacity scaling order w.r.t. $N$, ${P_{\mathrm{B}}}$ and ${P_{\mathrm{I}}}$, which is summarized in Table \ref{comp_R}. The term ``constant'' refers to the capacity achieved at a constant order, since the SNR is upper-bounded as the system parameters approach infinity.

From Table \ref{comp_R}, it is observed that BHU provides a higher capacity scaling order w.r.t. $N$ thanks to the passive beamforming gain, whereas the rate performance is limited by the AIRS under the schemes with double reflections. Moreover, we observe that the capacity of both BAPU and BPAU schemes does not increase with ${P_{\mathrm{B}}}$, whereas that of the BHU scheme linearly increases with ${P_{\mathrm{B}}}$. The results show that BHU is preferable provided that the total number of IRS elements or the transmit power is sufficiently large. 

It is observed from Table \ref{comp_R} that the capacity achieved by all the schemes remains constant as ${P_{\mathrm{I}}}$ changes, which indicates that further comparison is required. This is because the beamforming gain depends on the number of IRS elements but not on the amplification power. Moreover, although the desired signal power increases with ${P_{\mathrm{I}}}$, the power of the introduced noise also increases. In the large-${P_{\mathrm{I}}}$ regime, i.e., ${P_{\mathrm{I}}} \to \infty$, we have $\mathop {\lim }\limits_{{P_{\rm{I}}} \to \infty } \frac{{{R_0}}}{{{R_1}}} \to \frac{{{{\log }_2}( {1 + {P_{\rm{B}}}\beta _{{\rm{BI}}}^2{N_{\rm{a}}}/\sigma _{\rm{r}}^2} )}}{{{{\log }_2}( {1 + {P_{\rm{B}}}\beta _{\rm{B}}^2{N_{\rm{a}}}/\sigma _{\rm{r}}^2})}}$ and $\mathop {\lim }\limits_{{P_{\rm{I}}} \to \infty } \frac{{{R_0}}}{{{R_2}}} \to \frac{{{{\log }_2}( {1 + {P_{\rm{B}}}\beta _{{\rm{BI}}}^2{N_{\rm{a}}}/\sigma _{\rm{r}}^2} )}}{{{{\log }_2}( {1 + {P_{\rm{B}}}\beta _{\rm{B}}^2\beta _{\rm{I}}^2{N_{\rm{a}}}N_{{\rm{pas}}}^2/\sigma _{\rm{r}}^2} )}}$. Generally, the HIRS is placed further away from the BS than the IRS 1 in order to cover the user, which demonstrates that BAPU and BPAU are more appealing when a large amplification power is affordable. Specifically, BPAU achieves the best rate performance among the three schemes if $\beta _{\mathrm{I}}^2N_{{\mathrm{p}}}^2 > 1$ in the large $P_\mathrm{I}$ regime. Otherwise, BAPU performs best. 

\section{Numerical Results}
\label{Simulation}
In this section, numerical results are provided to verify our theoretical findings on the comparison of the three IRS deployment schemes. The path-loss exponents of all the links are set to 2. The path loss at a reference distance of 1 m is $-43$ dB. The other system parameters are set as follows: $P_\mathrm{B} = 20$ dBm, $P_\mathrm{I} = 8$ dBm, and $\sigma_0^2 = \sigma_{\mathrm{r}}^2 = -80$ dBm.

\subsection{IRS Elements Allocation Optimization}
In this subsection, we first validate the effectiveness of the proposed IRS elements allocation strategy. Then, we compare the performance of three IRS deployment schemes in terms of achievable rate. The BS, IRS 1, IRS 2, and user are located at (0,0,0) m, (5,0,5) m, (85,0,5) m, and (0,0,90) m, respectively. Given the same product distance for a fair comparison, we set the BS-HIRS distance as 80 m and the HIRS-user distance as 50 m. The following schemes are considered: 1) \textbf{Exh. search:} Exhaustive search method is adopted to determine the optimal elements allocation without the considered approximations; 2) \textbf{Equal-EA:} The number of active elements is equal to that of the passive elements, i.e., $N_\text{p}=N_\text{a}=N/2$.

\subsubsection{Accuracy of IRS Elements Allocation Design}
\begin{figure}[t]
	\centering
	\includegraphics[width=0.38\textwidth]{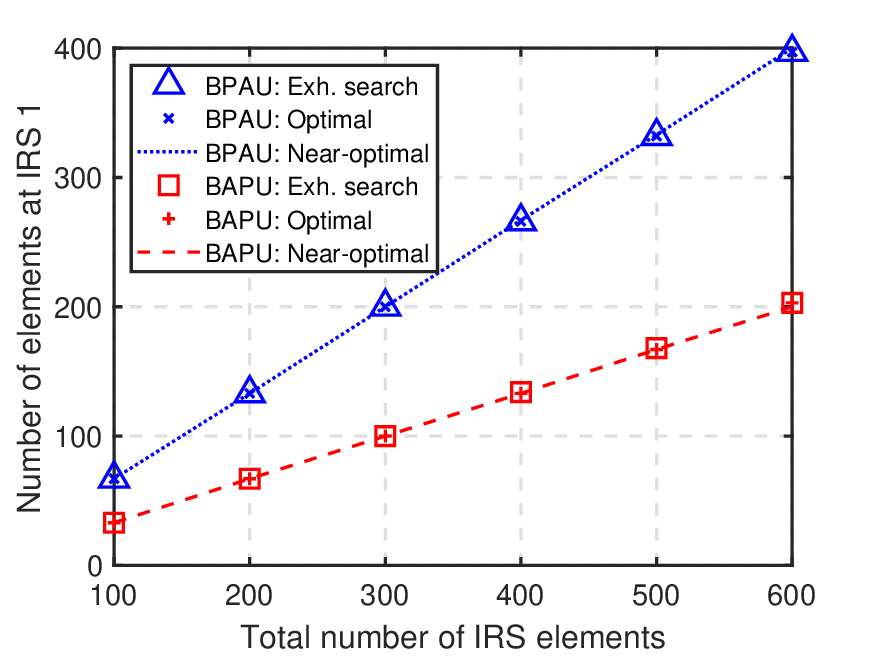}
	\vspace{-5pt}
	\caption{Number of elements at IRS 1 versus total number of elements $N$.}
	\label{fig:N_Np}
	\vspace{-5pt}
\end{figure}
In Fig. \ref{fig:N_Np}, we plot the number of elements at IRS 1 versus the total number of reflecting elements $N$. As shown in Fig. \ref{fig:N_Np}, the near-optimal solutions proposed in Section \ref{EAP_opt} closely match the results obtained by the exhaustive search method, thereby validating their effectiveness. When $N = 100$, the optimized elements are $N_1 = 33$ at IRS 1 and $N_2=67$ at IRS 2 for BAPU scheme, and $N_1=67$ and $N_2 = 33$ for BPAU scheme. The results show that more elements should be deployed at PIRS in both systems, benefiting from its high beamforming gain. When $N = 600$, $N_1=203$ and $N_2=397$ for the BAPU scheme, and $N_1=397$ and $N_2=203$ for BPAU scheme, which implies that the optimized elements for both IRSs increase with $N$. The results agree with the analysis of Proposition \ref{pro_N_AP} and \ref{pro_N_PA}.

\subsubsection{Impact of Total Number of IRS Elements}
\begin{figure}[t]
	\centering
	\includegraphics[width=0.38\textwidth]{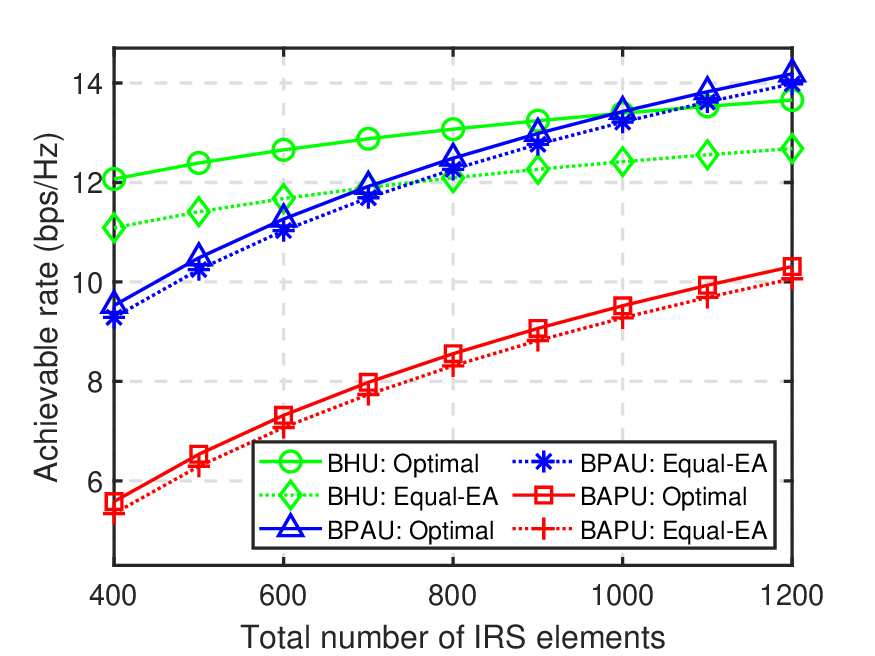}
	\vspace{-5pt}
	\caption{Achievable rate versus total number of elements $N$.}
	\label{fig:R_N_opt_N}
	\vspace{-5pt}
\end{figure}
To unveil the benefits of the proposed design, we study the impact of the total number of IRS elements $N$ on the achievable rate, by plotting it versus $N$ in Fig. \ref{fig:R_N_opt_N}. It is observed that the achievable rates of both BAPU and BPAU schemes grow faster than that of BHU, owing to their higher SNR scaling order (i.e., $\mathcal{O} (N^3)$ versus $\mathcal{O} (N^2)$). Moreover, we observe that the optimal IRS elements allocation design outperforms the equal elements allocation, which highlights the importance of determining the number of active and passive elements at the IRS, especially for the BHU scheme. Note that the PIRS should be placed close to the BS or user, while the AIRS prefers to be close to the user. BAPU performs much better than BPAU because it is deployed at an inappropriate location, which motivates us to investigate the impact of IRS placement for further comparison. 

\subsection{IRS Placement Optimization}
In this subsection, we first validate the effectiveness of the proposed IRS placement design. Subsequently, we shed light on the rate performance comparison of three IRS deployment schemes. The number of active and passive elements are set as $N_\mathrm{a} = 200$ and $N_\mathrm{p} = 500$, respectively. The height of IRS for BHU scheme is set as $h_\mathrm{s} = 10$ m, and that for BAPU and BPAU scheme is set as $h_\mathrm{d} = 5$ m. We consider the following schemes: 1) \textbf{Exh. search:} Exhaustive search method is adopted to find the optimal IRS placement without the considered approximations; 2) \textbf{Middle:} The IRS location is fixed where the horizontal distance between BS and AIRS/HIPS is set to $L/2$.

\subsubsection{Accuracy of IRS Placement Design}
\begin{figure}[t]
	\centering
	\includegraphics[width=0.38\textwidth]{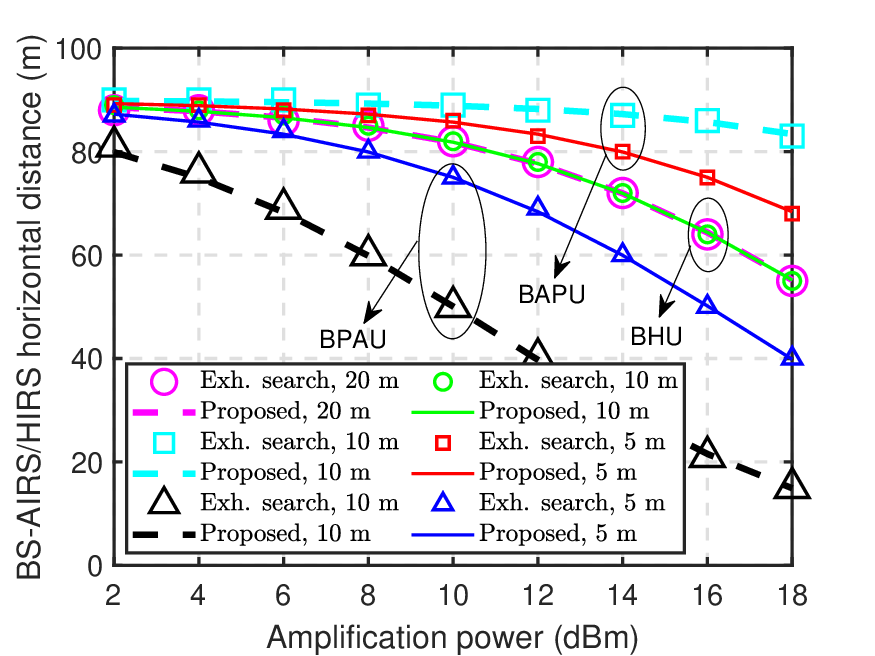}
	\vspace{-5pt}
	\caption{Optimized IRS placement versus amplification power $P_\mathrm{I}$.}
	\label{fig:x_Pi}
	\vspace{-5pt}
\end{figure}
In Fig. \ref{fig:x_Pi}, we plot the optimized horizontal distance between BS and AIRS/HIRS versus amplification power at different heights of IRSs. First, one can observe that our proposed near-optimal placement is close to the optimal one obtained by exhaustive search in the cases of different IRS heights, which validates its effectiveness. Second, we observe that the BS-AIRS and BS-HIRS horizontal distance decreases as $P_\mathrm{I}$ increases. With a larger $P_\mathrm{I}$, the AIRS/HIRS should be deployed closer to the BS to attenuate the channel gain, thereby reducing the noise amplification. Third, it is observed that the height of the IRS has a more significant effect on the optimized horizontal position under the double-reflection case than under the single-reflection case. Specifically, for the BAPU scheme, the AIRS prefers to be closer to the passive IRS to minimize the path loss. Moreover, the two distributed IRSs tend to be closer together for the BPAU scheme and farther apart for the BAPU scheme, which shows the importance of properly designing appropriate placement strategies for the schemes with different deployment orders of AIRS and PIRS.

\subsubsection{Impact of Amplification Power}
\begin{figure}[t]
	\centering
	\includegraphics[width=0.38\textwidth]{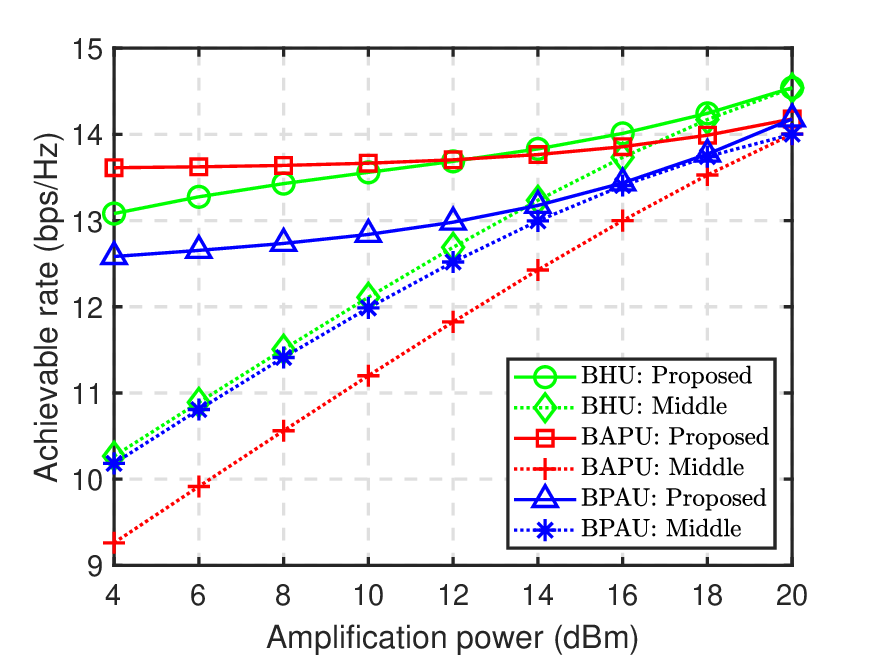}
	\vspace{-5pt}
	\caption{Achievable rate versus amplification power $P_\mathrm{I}$.}
	\label{fig:R_x_opt_Pi}
	\vspace{-5pt}
\end{figure}
To show the superiority of the proposed IRS placement design, we study the impact of the amplification power of active elements on the achievable rate, by plotting it versus the amplification power $P_\mathrm{I}$ in Fig. \ref{fig:R_x_opt_Pi}. First, one can observe that the rates achieved by all the schemes increase with the $P_\mathrm{I}$ due to the higher signal amplification gain. Second, it is observed that the proposed IRS placement design can significantly improve the achievable rate compared to the fixed placement strategies. The performance gap is pronounced when the amplification power is small. This is because it enjoys the advantages of dynamically striking a balance between signal and noise amplification at the active elements. Moreover, we observe that the BHU scheme outperforms the BPAU scheme over the whole considered $P_\mathrm{I}$ regime, which agrees with our analysis in \eqref{SNR_PA_H}. Furthermore, BHU performs best when $P_\mathrm{I} > 12$ dBm since it can fully exploit the beamforming and amplification gains provided by active elements. When $P_\mathrm{I} = P_\mathrm{B} = 20$ dBm, it is observed that BAPU and BPAU achieve the same rate performance, which validates our analysis in \eqref{con_x_ap_pa}.

\subsection{Performance Comparison via Joint Optimization}
For comparison, we consider the following benchmarks: \textbf{1) BPU \cite{singlePassive}:} The transmission link is BS→PIRS→user, where the PIRS is consists of $N$ passive elements and placed directly above the user; \textbf{2) BPPU \cite{doublePIRS_N}:} The transmission link is BS→PIRS 1→PIRS 2→user, where both PIRSs are deployed directly above the BS and user, respectively, each with $N/2$ passive elements. The height of IRSs are same as those for Fig. \eqref{fig:x_Pi} and Fig. \eqref{fig:R_x_opt_Pi}.
\subsubsection{Impact of Total Number of IRS Elements}
\begin{figure}[t]
	\centering
	\includegraphics[width=0.38\textwidth]{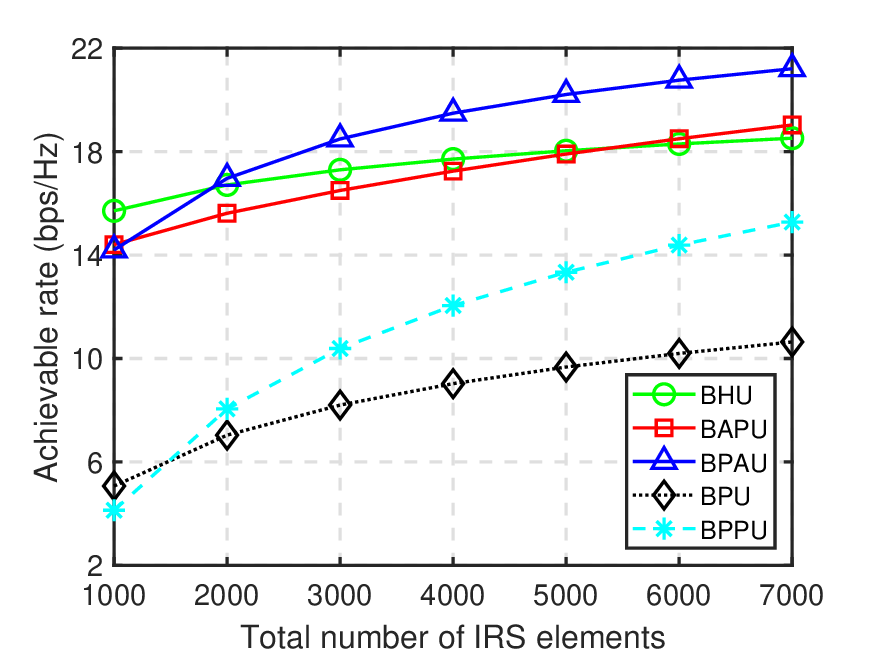}
	\vspace{-5pt}
	\caption{Achievable rate versus total number of elements $N$.}
	\label{fig:R_opt_N}
	\vspace{-5pt}
\end{figure}
In Fig. \ref{fig:R_opt_N}, the achievable rate versus the total number of IRS elements $N$ is plotted to unveil the impact of the elements budget on the system's performance. It is observed that BAPU and BPAU perform similarly and are both worse than BHU when the total number of IRS elements is small, which indicates that BHU is a preferred choice for the size-limited case. As $N$ increases, BPAU outperforms BAPU. Moreover, both schemes perform better than BHU because the SNR of systems with the first two schemes increase with $N$ as $\mathcal{O} (N^3)$ thanks to the IRS elements allocation optimization, whereas that with the latter increases with $N$ as $\mathcal{O} (N^2)$. The result emphasizes the significance of carefully optimizing the IRS elements allocation to fully exploit the increase in large $N$ for improving the performance of the systems with BAPU or BPAU schemes. Moreover, we observe that the three considered schemes outperform BPPU and BPU schemes when the total number of IRS elements is moderate, which unveils their effectiveness.  

\subsubsection{Impact of Amplification Power}
\begin{figure}[t]
	\centering
	\includegraphics[width=0.38\textwidth]{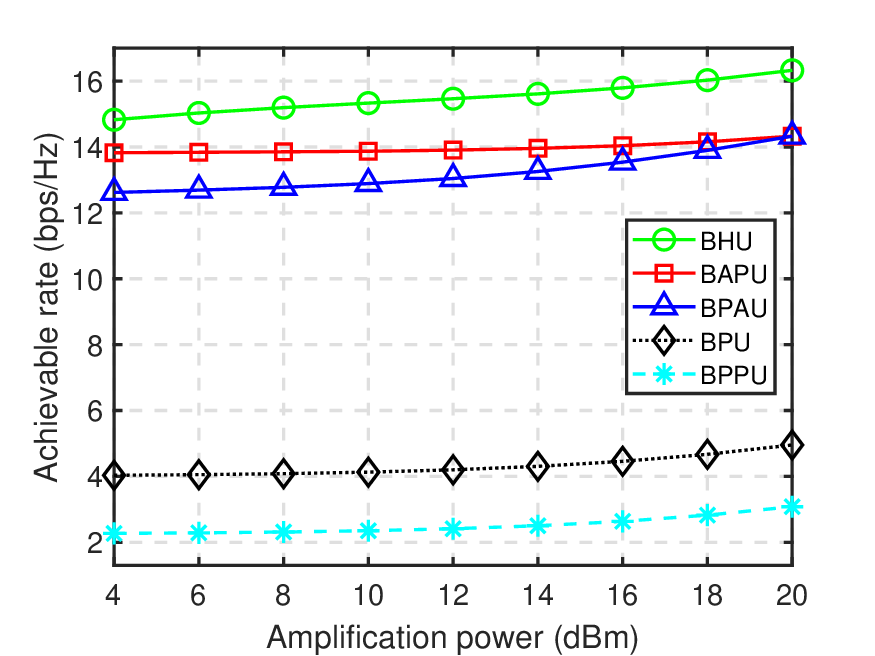}
	\vspace{-5pt}
	\caption{Achievable rate versus amplification power $P_\mathrm{I}$.}
	\label{fig:R_opt_Pi}
	\vspace{-5pt}
\end{figure}
In Fig. \ref{fig:R_opt_Pi}, we plot the achievable rate versus amplification power $P_\mathrm{I}$. For the entire range of the amplification power, the schemes with active elements outperform those with passive elements only because they provide a high amplification gain. Moreover, BHU achieves the best rate performance under the limited IRS element budget thanks to its optimal IRS elements allocation and placement, which strikes a good balance between the power amplification gain of active elements and the beamforming gain provided by the passive elements. Since the HIRS comprises both active and passive elements, the multiplicative path loss effect can be effectively alleviated via the flexible elements allocation compared to the single/double PIRS. 

Note that AIRS can effectively mitigate the multiplicative fading effect. When $P_\mathrm{I}$ is small, BAPU performs better than BPAU because it suffers less path loss over shorter signal propagation distance. However, the achievable rate of BAPU remains almost constant under different amplification power. It inspires us to focus on the configuration of other parameters rather than the amplification power when applying BAPU with optimal IRS elements allocation and placement. BAPU and BPAU achieve the same achievable rate when $P_\mathrm{I} = P_\mathrm{B}$, which also agrees with our theoretical analysis for the IRS placement optimization. It indicates that the performance bottleneck of BPAU mainly lies in the insufficient amplification power. As such, increasing the amplification coefficient at each element of AIRS by increasing the amplification power budget can compensate for the performance loss caused by the limited total number of IRS elements with the BPAU scheme. The results highlight the importance of the placement order of distributed AIRS and PIRS to compensate for the path loss.

\subsubsection{Impact of Transmit Power}
\begin{figure}[t]
	\centering
	\includegraphics[width=0.38\textwidth]{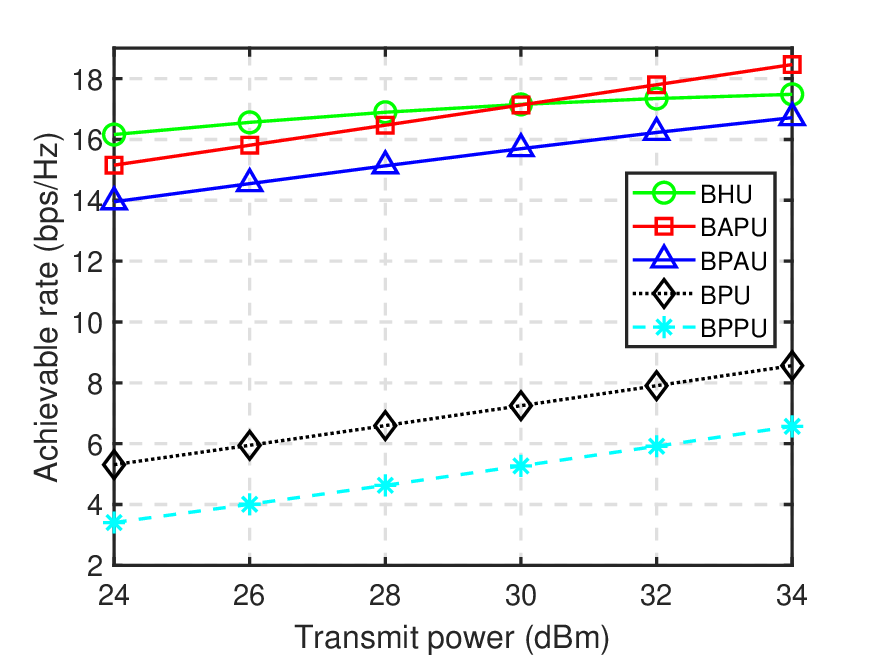}
	\vspace{-5pt}
	\caption{Achievable rate versus transmit power $P_\mathrm{B}$.}
	\label{fig:R_opt_Pb}
	\vspace{-5pt}
\end{figure}
In Fig. \ref{fig:R_opt_Pb}, we investigate the impact of transmit power on achievable rate by plotting it versus transmit power $P_\mathrm{B}$. One can observe that BHU performs better than other schemes when $P_\mathrm{B} < 30$ dBm. It demonstrates the superiority of the HIRS and implies that the deployment of HIRS is beneficial for practical scenarios with a limited budget for the total number of IRS elements, amplification power, and transmit power. As $P_\mathrm{B}$ increases, the achievable rates of all the schemes increase because of the increasing beamforming gain brought by the IRS. However, the increase in transmit power leads to a decrease in the reflection coefficient, which severely limits the performance of the single-reflection link involving active elements in an HIRS-aided communication system. This is also the reason why the corresponding curve tends to converge when the transmission power is large. Compared to increasing the amplification power, increasing the transmission power has a greater effect on enhancing the rate performance, especially for the BAPU scheme. When $P_\mathrm{B} > 30$ dBm, we observe that the BAPU becomes the best choice for achievable rate maximization. This is because it not only provides a higher beamforming gain through double reflections but also enables a flexible balance between signal and noise amplification at the AIRS. The results highlight the importance of appropriately configuring the transmit power for BAPU to fully harness its potential for enhancing rate performance.

\section{Conclusion}
\label{Conclusion}
In this paper, we studied the capacity of the IRS-aided communication system with both active and passive elements. In particular, three deployment schemes were considered, namely, BHU, BAPU, and BPAU schemes. The optimized elements allocation and placement design for the three schemes was provided, which lays the foundation for theoretical performance comparison. Our analysis revealed that more passive elements should be deployed under the BAPU and BPAU schemes. Moreover, the HIRS/AIRS should be placed closer to the BS as the amplification power increases. With the proposed joint design of the IRS elements allocation and placement, the BHU scheme can be a practically promising solution for maximizing the capacity and enabling energy-efficient communication in space-limited scenarios. In contrast, BAPU achieves better rate performance with sufficient transmit power, whereas BPAU is preferable for a large total number of IRS elements. Numerical results validated the theoretical findings and demonstrated that our proposed systems outperform other benchmark systems in terms of achievable rate under different system setups.

To facilitate the derivation of IRS deployment for rate performance comparison under different transmission schemes, we assume that the BS is equipped with a single antenna and the transmit power is fixed for tractability, as in \cite{HIRS_SE,CSI,single-antenna}. The results in this paper can be extended to the multi-antenna BS case in principle, however, dedicated algorithms are needed to optimize the transmit power/beamformer \cite{R1,multiPassive_x2}, which is left for our future work. With a higher transmit beamforming gain, it might be the case that more passive elements should be allocated at the HIRS to capture a higher passive beamforming gain, whereas more active elements should be deployed to increase the amplification gain under the BAPU/BPAU scheme. Moreover, it would be likely that the HIRS/AIRS moves closer to the user. By jointly designing the elements allocation and placement, the BAPU scheme may outperform the other two schemes. In addition, this paper can be extended in several promising directions to motivate future work. For example, the real effects of multi-path wireless environment can be more accurately characterized by considering the non-LoS paths, which is beneficial for further improving the IRS deployment design. Furthermore, it is worth investigating the impact of IRSs in the scenario with direct BS-user link. 

\section*{Appendix A: Proof of Proposition \ref{pro_N_H}}
Note that ${\gamma _0}$ is a function w.r.t. $\tilde N_\mathrm{a}$ and thus we use $\gamma_\mathrm{h} ( {{{\tilde N}_{{\mathrm{a}}}}} ) = {c_1}{\left( {{c_2}\sqrt {{\tilde N_{{\mathrm{a}}}}}  + {N_{{\mathrm{p}}}}} \right)^2}$ to represent ${\gamma _0}$. The first-order partial derivative of $\gamma_\mathrm{h} ( {{{\tilde N}_{{\mathrm{a}}}}} )$ w.r.t. $\sqrt{\tilde N_\mathrm{a}}$ is given by
\begin{align}
	\label{first_H}
	\frac{{\partial \gamma_\mathrm{h} ( {{{\tilde N}_{{\mathrm{a}}}}} )}}{{\partial \sqrt {{{\tilde N}_{{\mathrm{a}}}}} }} \!= 2{c_1}( { - {{\tilde N}_{{\mathrm{a}}}} + {c_2}\sqrt {{{\tilde N}_{{\mathrm{a}}}}}  + N} )( { - 2\sqrt {{{\tilde N}_{{\mathrm{a}}}}}  + {c_2}} ).
\end{align}
Note that ${c_1} > 0$. Moreover, we have ${ - {{\tilde N}_{{\mathrm{a}}}} + {c_2}\sqrt {{{\tilde N}_{{\mathrm{a}}}}}  + N} > 0, \forall \sqrt{\tilde N_\mathrm{a}}$ since $\alpha_{\mathrm{0}} \ge 1$, i.e., $c_2 \ge \sqrt{\tilde N_\mathrm{a}}$. Thus, the roots for \eqref{first_H} are give by $\sqrt{\tilde N_\mathrm{a}} = x_{\mathrm{0,rt}}  \buildrel \Delta \over = {c_2}/2$. When $\sqrt{\tilde N_\mathrm{a}} \le x_{\mathrm{0,rt}}$, it follows that $\frac{{\partial \gamma_\mathrm{h} ( {{{\tilde N}_{{\mathrm{a}}}}} )}}{{\partial \sqrt {{{\tilde N}_{{\mathrm{a}}}}} }} > 0 $, i.e., $\gamma_\mathrm{h} ( {{{\tilde N}_{{\mathrm{a}}}}} )$ monotonically increases with ${{{\tilde N}_{{\mathrm{a}}}}}$. Thus, $\gamma_\mathrm{h} ( {{{\tilde N}_{{\mathrm{a}}}}} )$ is maximized at $\tilde N_\mathrm{a} = N - 1$. Otherwise, it follows that $\frac{{\partial \gamma_\mathrm{h} ( {{{\tilde N}_{{\mathrm{a}}}}} )}}{{\partial \sqrt {{{\tilde N}_{{\mathrm{a}}}}} }} > 0, \forall {{{\tilde N}_{{\mathrm{a}}}}} \in (0, x_{\mathrm{0,rt}}^2)$ and $\frac{{\partial \gamma_\mathrm{h} ( {{{\tilde N}_{{\mathrm{a}}}}} )}}{{\partial \sqrt {{{\tilde N}_{{\mathrm{a}}}}} }} < 0, \forall {{{\tilde N}_{{\mathrm{a}}}}} \in (x_{\mathrm{0,rt}}^2, N )$, which implies that $\gamma_\mathrm{h} ( {{{\tilde N}_{{\mathrm{a}}}}} )$ first increases and then decreases with ${{{\tilde N}_{{\mathrm{a}}}}}$. Thus, $\gamma_\mathrm{h} ( {{{\tilde N}_{{\mathrm{a}}}}} )$ is maximized at $\tilde N_\mathrm{a} = x_{\mathrm{0,rt}}^2$. Based on the above results, we complete the proof.

\section*{Appendix B: Proof of Proposition \ref{pro_N_AP}}
Note that ${\gamma _1}$ is a function w.r.t. $\tilde N_\mathrm{p}$ and thus we use $\gamma_\mathrm{ap} ( {{{\tilde N}_{{\mathrm{p}}}}} ) = \frac{{{c_6}{N_{{\mathrm{a}}}}\tilde N_{{\mathrm{p}}}^2}}{{{c_3}\tilde N_{{\mathrm{p}}}^2 + {c_4}}}$ to represent ${\gamma _1}$, where the first-order partial derivative of $\gamma_\mathrm{ap} ( {{{\tilde N}_{{\mathrm{p}}}}} )$ w.r.t. $\tilde N_\mathrm{p}$ is given by
\begin{align}
	\label{first_AP}
	\frac{{\partial {\gamma_\mathrm{ap} ( {{{\tilde N}_{{\mathrm{p}}}}} )}}}{{\partial {{\tilde N}_{{\mathrm{p}}}}}} = \frac{{{c_6}{{\tilde N}_{{\mathrm{p}}}}( { - {c_3}\tilde N_{{\mathrm{p}}}^3 - 3{c_4}{{\tilde N}_{{\mathrm{p}}}} + 2N{c_4}} )}}{{{{( {{c_3}\tilde N_{{\mathrm{p}}}^2 + {c_4}} )^2}}}}.
\end{align}
Let $x = {\tilde N}_{{\mathrm{p}}}$ and $g_1\left(x\right) \buildrel \Delta \over = { - {c_3}x^3 - 3{c_4}x + 2N{c_4}}$. The first-order partial derivative of $g_1\left(x\right)$ w.r.t. $x$ is given by $g_1'\left( x \right) = -3{c_3}x^2 - 3{c_4}$. Since $g_1'\left( x \right) < 0$ for $x \in \left[ 0, N \right]$, $g_1\left(x\right)$ monotonically decreases with $x$ for $x \in \left[ 0, N\right]$. Hence, the single unique root of $g_1\left(x\right) = 0$ is denoted by $x_1^\mathrm{rt}$. When $x \in [ 0, x_1^\mathrm{rt} )$, we have $g_1\left( x \right) > 0$ and $\frac{{\partial {\gamma_\mathrm{ap} ( {{{\tilde N}_{{\mathrm{p}}}}} )}}}{{\partial {{\tilde N}_{{\mathrm{p}}}}}} > 0$, which implies that $\gamma_\mathrm{ap} ( {{{\tilde N}_{{\mathrm{p}}}}} )$ monotonically increases with ${\tilde N}_{{\mathrm{p}}}$. When $x \in ( x_1^\mathrm{rt}, N ]$, it follows that $g_1\left(x\right) < 0$ and $\frac{{\partial {\gamma_\mathrm{ap} ( {{{\tilde N}_{{\mathrm{p}}}}} )}}}{{\partial {{\tilde N}_{{\mathrm{p}}}}}} < 0$, i.e., $\gamma_\mathrm{ap} ( {{{\tilde N}_{{\mathrm{p}}}}} )$ monotonically decreases with ${\tilde N}_{{\mathrm{p}}}$. Accordingly, $\gamma_\mathrm{ap} ( {{{\tilde N}_{{\mathrm{p}}}}} )$ is maximized when ${\tilde N}_{{\mathrm{p}}} = x_1^\mathrm{rt}$. Proposition \ref{pro_N_AP} is thus proved.

\section*{Appendix C: Proof of Proposition \ref{pro_x}}
The first-order partial derivative of ${\gamma _{{\mathrm{0,x}}}} (x_\mathrm{BI})$ w.r.t. $x_\mathrm{BI}$ is given by
\begin{align}
	\frac{{\partial {\gamma _{{\mathrm{0,x}}}} (\! {{x_{{\mathrm{BI}}}}} \!)}}{{\partial {x_{{\mathrm{BI}}}}}} \!\!=\!\! \frac{{ - 2N_{{\mathrm{p}}}^2P_{\mathrm{B}}^3{\beta ^3}{x_{{\mathrm{BI}}}}{{( {\sqrt {{N_{{\mathrm{a}}}}{P_{\mathrm{I}}}}  \!\!+\!\! {f_1}(\! {{x_{{\mathrm{BI}}}}} \!)} )^3}}{f_2}(\! {{x_{{\mathrm{BI}}}}} \!)}}{{{{ ( {x_{{\mathrm{BI}}}^2 + h_\mathrm{s}^2} )^2}}{f_1}( {{x_{{\mathrm{BI}}}}})f_3^2 ( {{x_{{\mathrm{BI}}}}} )}},
\end{align}
where ${f_1}\left( {{x_{{\mathrm{BI}}}}} \right) = \sqrt {N_{{\mathrm{p}}}^2{P_{\mathrm{B}}}\beta /\left( {x_{{\mathrm{BI}}}^2 + h_\mathrm{s}^2} \right)} $,
\begin{align}
	{f_2}\left( {{x_{{\mathrm{BI}}}}} \right) =& \frac{{{{\left( {x_{{\mathrm{BI}}}^2 + h_\mathrm{s}^2} \right)^2}}{f_1}\left( {{x_{{\mathrm{BI}}}}} \right)\left( {{c_{10}}{x_{{\mathrm{BI}}}} - {c_9}\left( {L - {x_{{\mathrm{BI}}}}} \right)} \right)}}{{N_{{\mathrm{p}}}^2P_{\mathrm{B}}^2{\beta ^2}{x_{{\mathrm{BI}}}}\left( {\sqrt {{N_{{\mathrm{a}}}}{P_{\mathrm{I}}}}  + {f_1}\left( {{x_{{\mathrm{BI}}}}} \right)} \right)}} \nonumber\\
	&+ \frac{{{f_3}\left( {{x_{{\mathrm{BI}}}}} \right)}}{{{P_{\mathrm{B}}}\beta {{\left( {\sqrt {{N_{{\mathrm{a}}}}{P_{\mathrm{I}}}}  + {f_1}\left( {{x_{{\mathrm{BI}}}}} \right)} \right)^2}}}},
\end{align}
${f_3}\left( {{x_{{\mathrm{BI}}}}} \right) = {c_{10}}\left( {x_{{\mathrm{BI}}}^2 + h_\mathrm{s}^2} \right) + {c_9}\left( {{{\left( {L - {x_{{\mathrm{BI}}}}} \right)^2}} + h_\mathrm{s}^2} \right)$. In the high SNR case, i.e., ${{{P_{\mathrm{B}}}\beta {{\left( {\sqrt {{N_{{\mathrm{a}}}}{P_{\mathrm{I}}}}  + {f_1}\left( {{x_{{\mathrm{BI}}}}} \right)} \right)^2}}}}/{{{f_3}\left( {{x_{{\mathrm{BI}}}}} \right)}} \gg 1$, we have ${f_3}\left( {{x_{{\mathrm{BI}}}}} \right)/\left( {{P_{\mathrm{B}}}\beta {{\left( {\sqrt {{N_{{\mathrm{a}}}}{P_{\mathrm{I}}}}  + {f_1}\left( {{x_{{\mathrm{BI}}}}} \right)} \right)^2}}} \right) \ll 1$, which can be approximated as 0. Hence, the single unique root of ${f_2}\left( {{x_{{\mathrm{BI}}}}} \right)$ is denoted by $x_{{\mathrm{BI}}}^\mathrm{rt} = c_9 L/(c_9+c_{10})$. When $x_{{\mathrm{BI}}} \in [0, x_{{\mathrm{BI}}}^\mathrm{rt})$, we have ${f_2}\left( {{x_{{\mathrm{BI}}}}} \right) < 0$ and $\frac{{\partial {\gamma _{{\mathrm{0,x}}}} ( {{x_{{\mathrm{BI}}}}} )}}{{\partial {x_{{\mathrm{BI}}}}}} > 0$, which implies that ${{\gamma _{{\mathrm{0,x}}}}\left( {{x_{{\mathrm{BI}}}}} \right)}$ monotonically increases with $x_{{\mathrm{BI}}}$. When $x_{{\mathrm{BI}}} \in (x_{{\mathrm{BI}}}^\mathrm{rt}, L]$, it follows that ${f_2}\left( {{x_{{\mathrm{BI}}}}} \right) > 0$ and $\frac{{\partial {\gamma _{{\mathrm{0,x}}}} ( {{x_{{\mathrm{BI}}}}} )}}{{\partial {x_{{\mathrm{BI}}}}}} < 0$, i.e., ${{\gamma _{{\mathrm{0,x}}}}\left( {{x_{{\mathrm{BI}}}}} \right)}$ monotonically decreases with $x_{{\mathrm{BI}}}$. Thus, ${{\gamma _{{\mathrm{0,x}}}}\left( {{x_{{\mathrm{BI}}}}} \right)}$ is maximized at $x_{{\mathrm{BI}}} = x_{{\mathrm{BI}}}^\mathrm{rt}$, which completes the proof.

\bibliographystyle{IEEEtran}
\bibliography{refs.bib} 

\end{document}